\documentclass{aa6}

\usepackage{graphicx}
\usepackage{natbib}

\newcommand{\id}{\mbox{$\mathrm{d^{-1}}$}}
\newcommand{\Msun}{\mbox{$M_{\odot}$}}
\newcommand{\kms}{\mbox{$\mathrm{km\,s^{-1}}$}}

\newcommand{\T}{\mbox{$T_\mathrm{0}$}}

\newcommand{\Porb}{\mbox{$P_{\rm orb}$}}
\newcommand{\Psh}{\mbox{$P_{\rm sh}$}}
\newcommand{\Line}[3]{\Ion{#1}{#2}\,\,$\lambda$#3}

\newcommand{\Idline}[2]{\Ion{#1}{#2}}
\newcommand{\Ion}[2]{#1{\,\scriptsize #2}}
\newcommand{\Ha}{\mbox{${\mathrm H\alpha}$}}
\newcommand{\Hb}{\mbox{${\mathrm H\beta}$}}
\newcommand{\Hg}{\mbox{${\mathrm H\gamma}$}}

\newcommand{\msy}{\mbox{$\mathrm{\Msun\,yr^{-1}}$}}

\begin{document}

\title{Dwarf novae in the Hamburg Quasar Survey: Rarer than expected}

\author{A. Aungwerojwit\inst{1}\and 
        B.T. G\"ansicke\inst{1}\and
        P. Rodr\'iguez-Gil\inst{1,2}\and
        H.-J. Hagen\inst{3}\and
        S. Araujo-Betancor\inst{2}\and
	O. Baernbantner\inst{4}\and
        D. Engels\inst{3}\and
        R.E. Fried\inst{5}\and
        E.T. Harlaftis\inst{6}\and
        D. Mislis\inst{7}\and
	D. Nogami\inst{8}\and
	P. Schmeer\inst{9}\and
	R. Schwarz\inst{10}\and
	A. Staude\inst{10} \and
	M.A.P. Torres\inst{11}
        }
\authorrunning{Aungwerojwit et al.}
\titlerunning{Dwarf novae in the Hamburg Quasar Survey: Rarer than expected}

\offprints{A. Aungwerojwit, \\ e-mail: A.Aungwerojwit@warwick.ac.uk}

\institute{
   Department of Physics, University of Warwick, Coventry CV4 7AL, UK
\and
   Instituto de Astrof\'isica de Canarias, 38200 La Laguna, Tenerife, Spain
\and
   Hamburger Sternwarte, Universit\"at Hamburg, Gojenbergsweg
   112, 21029 Hamburg, Germany
\and 
  Universit\"ats-Sternwarte, Scheinerstr. 1, 81679 M\"unchen, Germany
\and
   Braeside Observatory, PO Box 906, Flagstaff AZ 86002, USA
\and
   \mbox{Institute of Space Applications and Remote Sensing,
   National Observatory of Athens, P.O. Box 20048, Athens 11810, Greece}
\and
   Department of Physics, Section of Astrophysics, Astronomy \&
   Mechanics, University of Thessaloniki, 541 24 Thessaloniki, Greece
\and
   Hida Observatory, Kyoto University, Kamitakara, Gifu 506-1314, Japan
\and
   Bischmisheim, Am Probstbaum 10, 66132 Saarbr\"ucken, Germany
\and
   Astrophysikalisches Institut Potsdam, An der Sternwarte 16, 14482
   Potsdam, Germany 
\and
   Harvard-Smithsonian Center for Astrophysics, 60 Garden St,
   Cambridge, MA 02138, USA
}

\date{Received \underline{\hskip2cm} ; accepted }

\abstract{}
{We report the discovery of five new dwarf novae that were
  spectroscopically identified in the Hamburg Quasar Survey (HQS), and
  discuss the properties of the sample of new dwarf novae from the
  HQS.}
{Follow-up time-resolved spectroscopy and photometry have been obtained
to characterise the new systems.}
{The orbital periods determined from analyses of the radial velocity
variations and/or orbital photometric variability are
$\Porb\simeq105.1$\,min or $\Porb\simeq109.9$\,min for HS\,0417+7445,
$\Porb=114.3\pm2.7$\,min for HS\,1016+3412, $\Porb=92.66\pm0.17$\,min
for HS\,1340+1524, $\Porb=272.317\pm0.001$\,min for HS\,1857+7127, and
$\Porb=258.02\pm0.56$\,min for HS\,2214+2845.  HS\,1857+7127 is found
to be partially eclipsing. In HS\,2214+2845 the secondary star of
spectral type M$3\pm1$ is clearly detected, and we estimate the
distance to the system to be $d=390\pm40$\,pc. We recorded one
superoutburst of HS\,0417+7445, identifying the system as a
SU\,UMa-type dwarf nova. HS\,1016+3412 and HS\,1340+1524 have rare
outbursts, and their subtype is yet undetermined. HS\,1857+7127 frequently
varies in brightness and may be a Z\,Cam-type dwarf
nova. HS\,2214+2845 is a {U\,Gem-type} dwarf nova with a most likely
cycle length of 71\,d.}
{{To date, 14 new dwarf novae have been identified in the HQS. The
ratio of short-period ($<3$\,h) to long-period ($>3$\,h) systems of
this sample is $1.3$, much smaller compared to the ratio of $2.7$
found for all known dwarf novae. The HQS dwarf novae display typically
infrequent or low-amplitude outburst activity, underlining the
strength of spectroscopic selection in identifying new CVs
independently of their variability. The spectroscopic
properties of short-period CVs in the HQS, newly identified and
previously known, suggest that most, or possibly all of them are
still evolving towards the minimum period. Their total number agrees
with the predictions of population models within an order of
magnitude. However, the bulk of all CVs is predicted to have evolved
past the minimum period, and those systems remain unidentified.  This
suggests that those post-bounce systems have markedly weaker H$\beta$
emission lines compared to the average known short-period CVs, and
undergo no or extremely rare outbursts.}}

\maketitle

\begin{figure*}
\includegraphics[width=5.8cm]{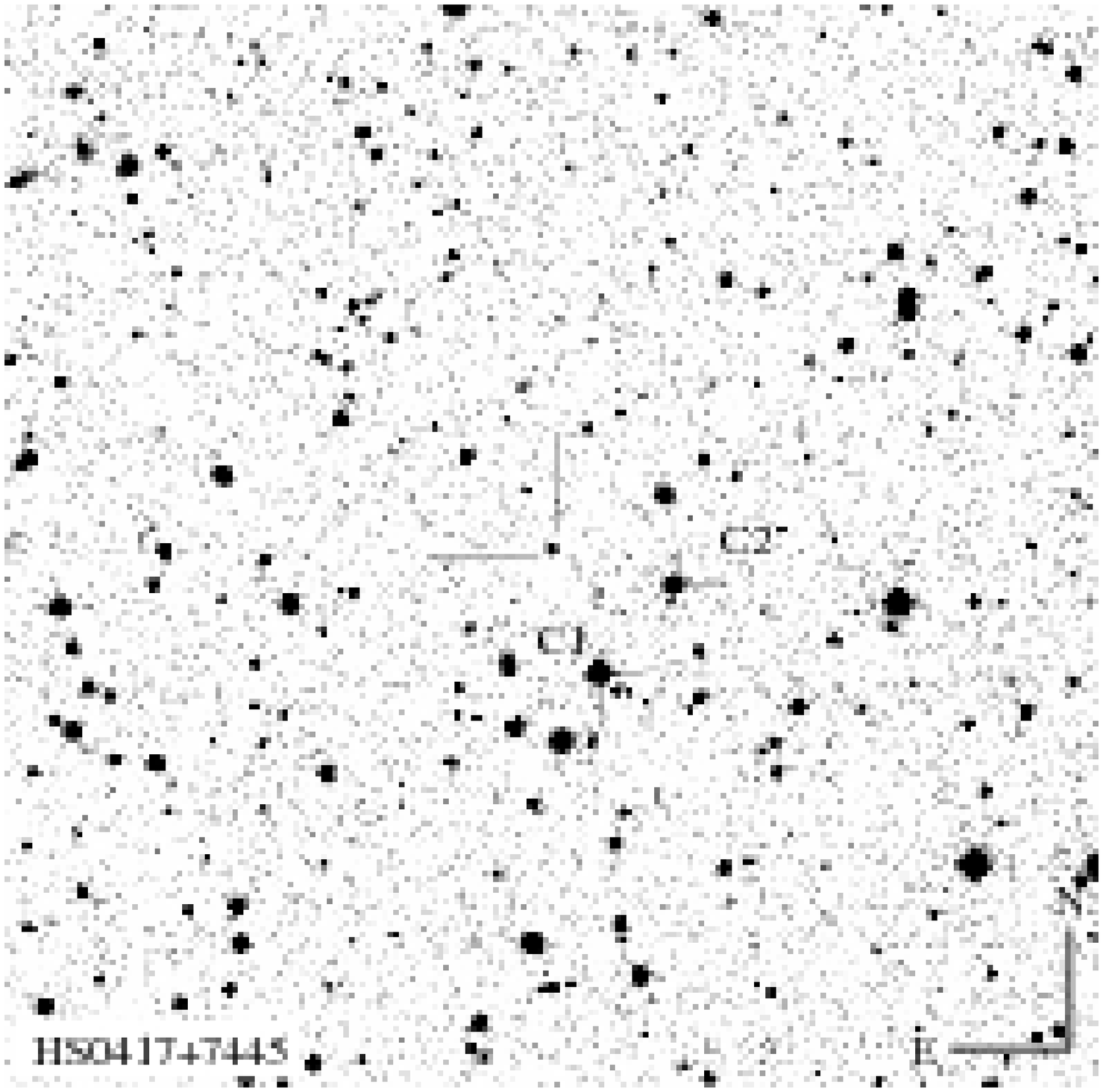}
\hfill
\includegraphics[width=5.8cm]{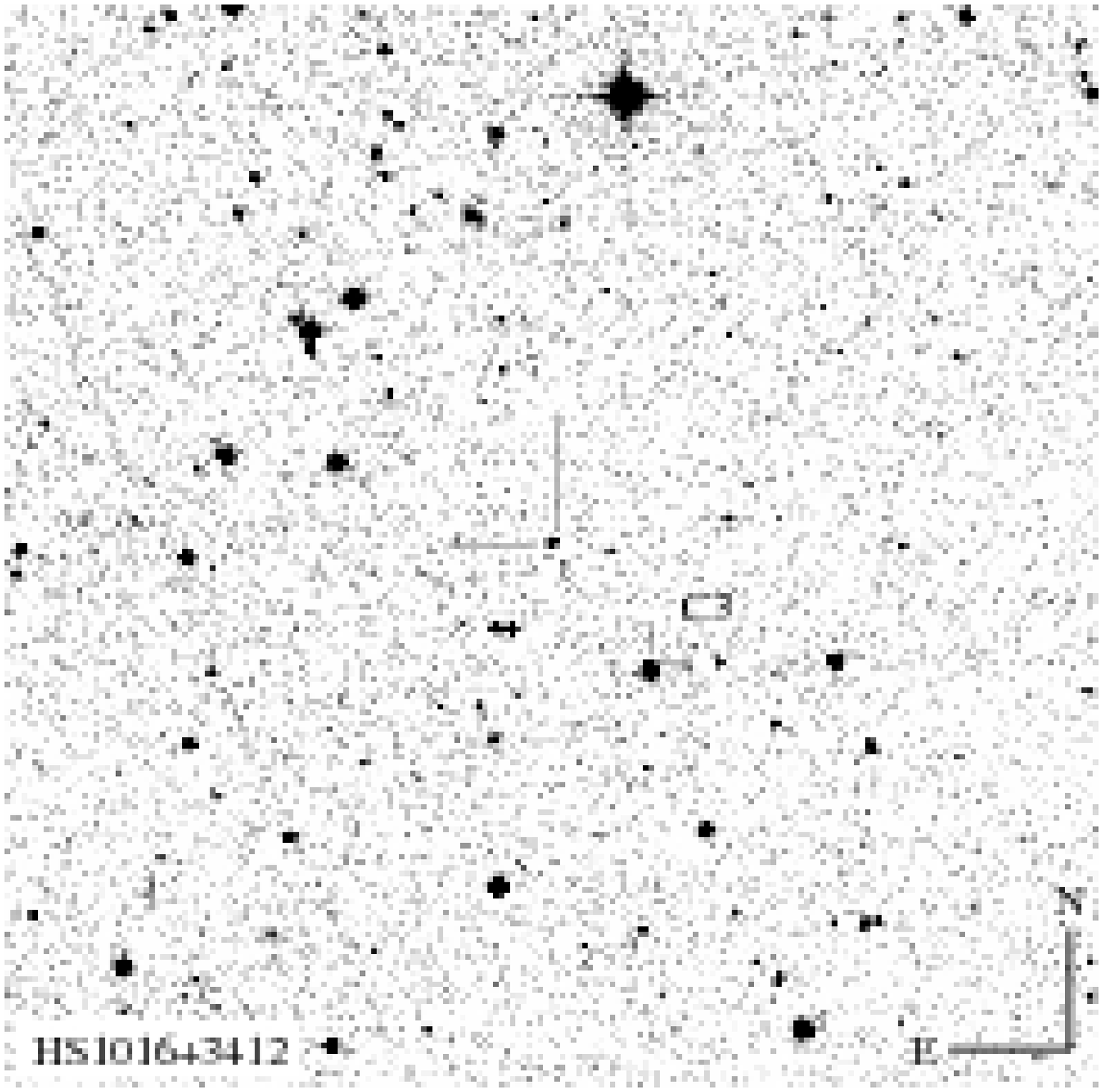}
\hfill
\includegraphics[width=5.8cm]{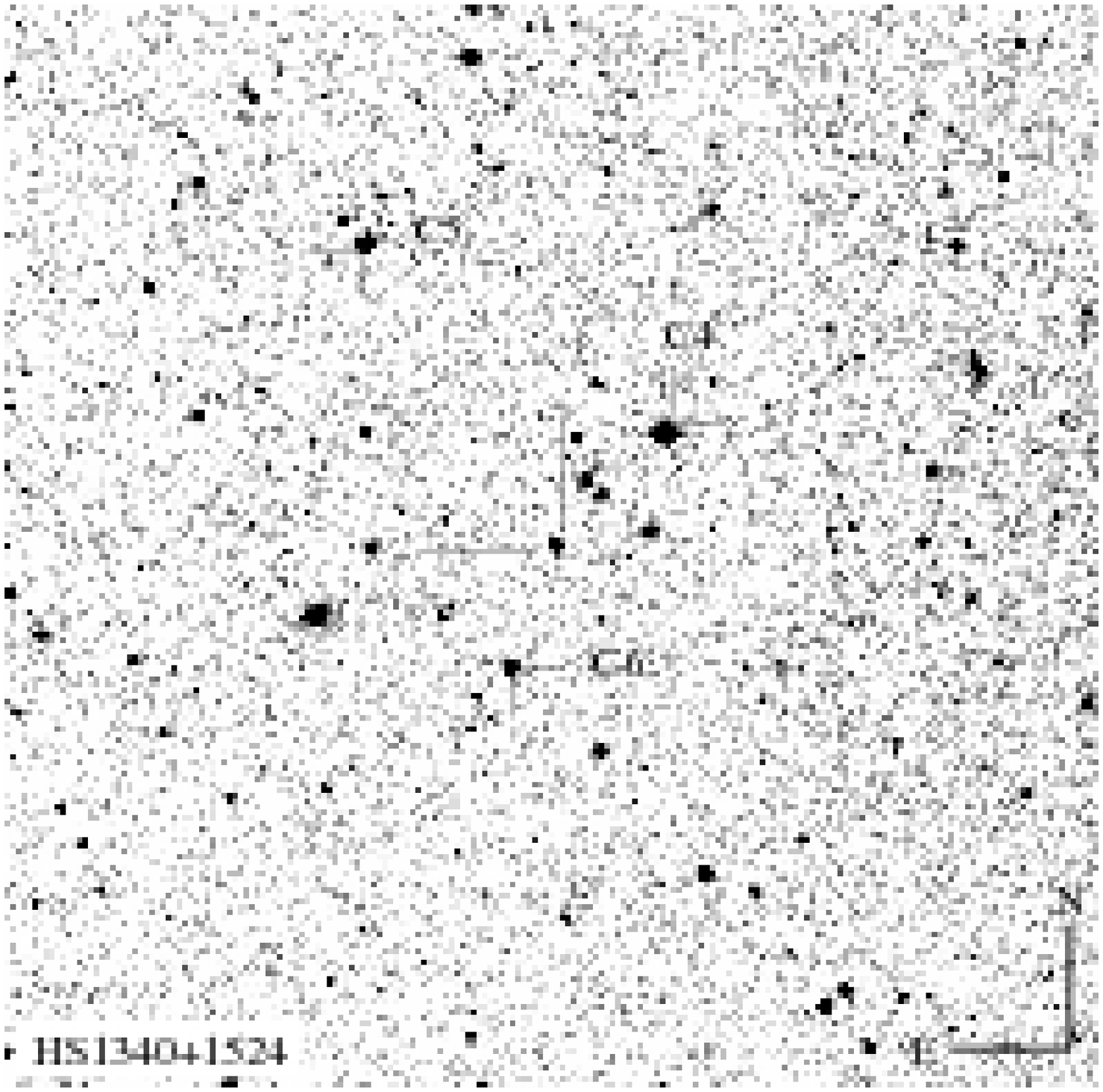}
\hfill
\includegraphics[width=5.8cm]{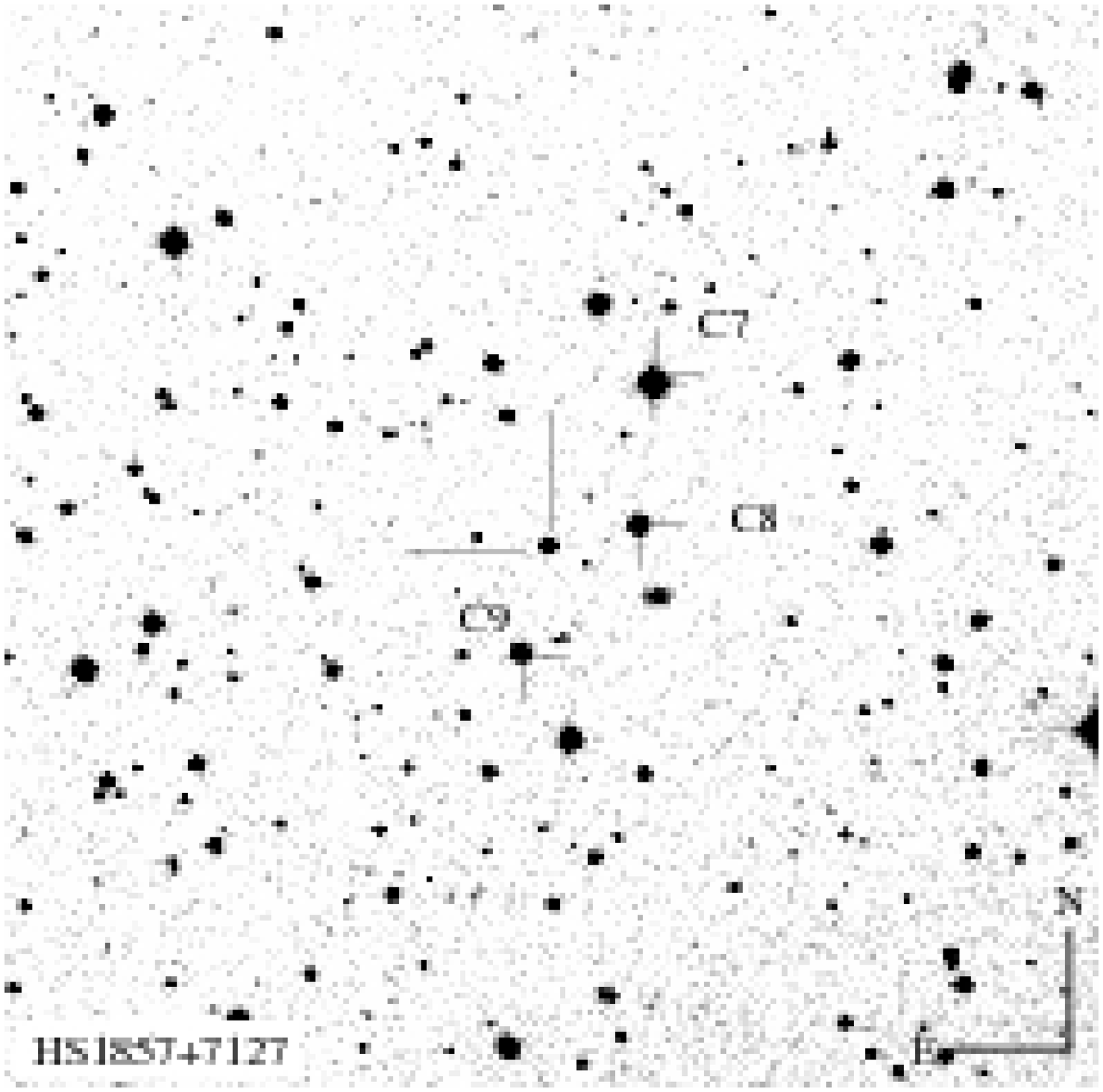}
\hfill
\includegraphics[width=5.8cm]{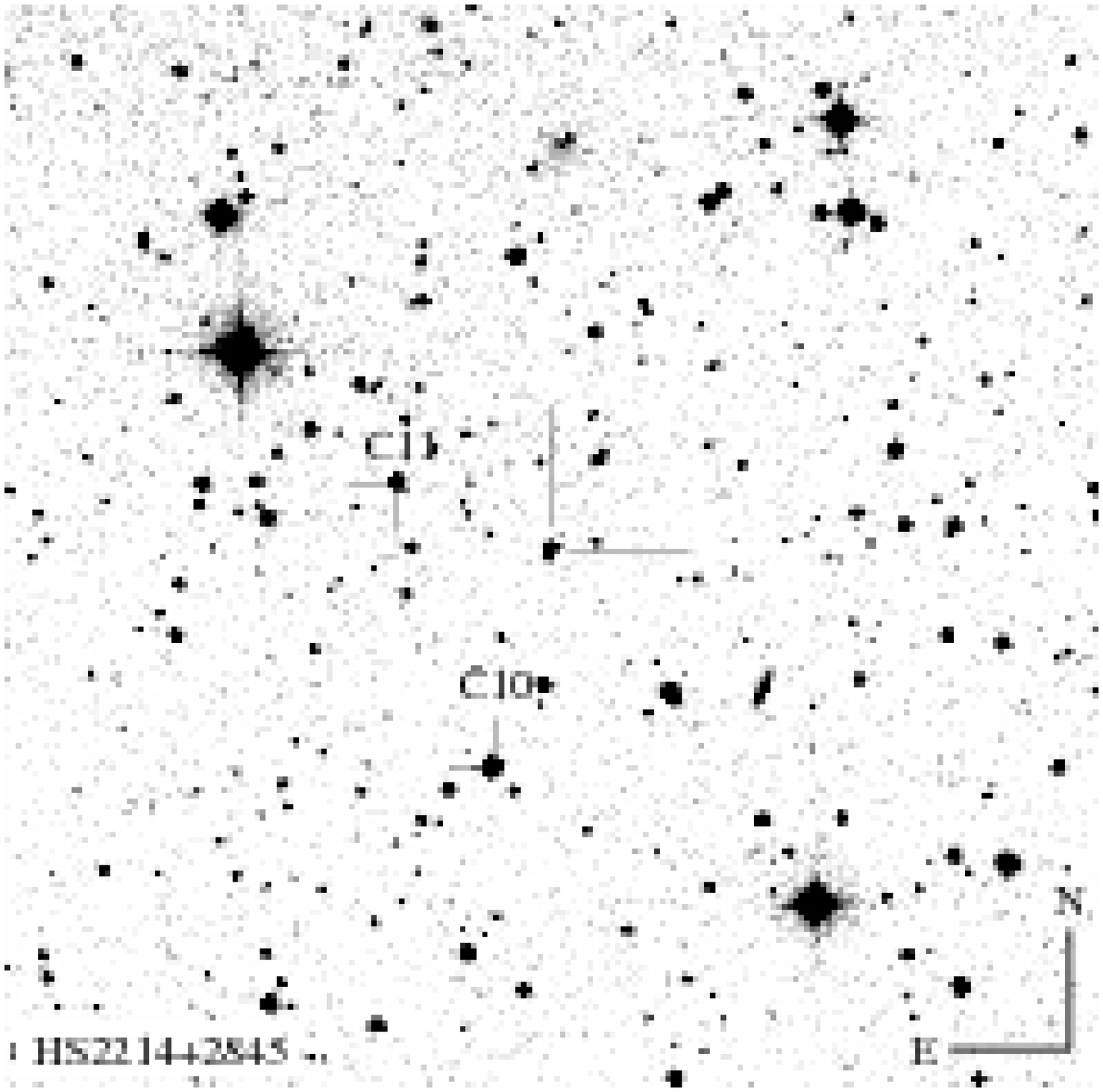}
\caption{\label{f-fc} $10\arcmin\times10\arcmin$ finding charts for
HS\,0417, HS\,1016, HS\,1340, HS\,1857, and HS\,2214
obtained from the Digitized Sky Survey.  HS\,2214 is the
northern component of a close visual binary. See
Table\,\ref{t-compstars} for details on the comparison stars C1--C11.
}
\end{figure*}

\section{Introduction}
Dwarf novae are a subset of non- (or weakly) magnetic cataclysmic
variables (CVs) that quasi-periodically brighten by several
magnitudes. The commonly accepted cause for dwarf nova outbursts is a
thermal instability in the accretion disc 
\citep[e.g.][]{cannizzoetal86-1, osaki96-1}. Within this theoretical
framework, accretion discs undergo outbursts if the accretion rate is
below a critical value, $\dot M_\mathrm{crit}$. Above the $2-3$\,h
period gap, the mass transfer rate in CVs is typically larger than
$\dot M_\mathrm{crit}$, and consequently only $\sim36$\,\% of
non-magnetic CVs with $\Porb>3$\,h are dwarf novae. For $\Porb<3$\,h,
the fraction of confirmed dwarf novae among the short-period
non-magnetic CVs is $\sim83$\,\%. The large difference in mass
transfer rate between short and long period systems is explained
within the standard CV evolution theory by the cessation of magnetic
braking~--~the dominant angular momentum loss mechanism in long period
CVs~--~ once the systems evolve down to 3\,h. Below $\Porb=3$\,h, the
evolution of CVs proceeds much slower due to lower mass transfer rates
driven by gravitational radiation as the angular momentum loss
mechanism \citep[e.g.][]{rappaportetal83-1, spruit+ritter83-1,
king88-1}. For such low mass transfer rates, the disc
instability model predicts thermally unstable accretion discs that
produce dwarf-nova outbursts\citep{cannizzo93-1,osaki96-1}.
Because of the long evolution time scale of low-mass transfer
CVs population models built upon the disrupted magnetic braking
scenario predict that 99\% of all CVs should have periods below $3$\,h
\citep{kolb93-1,howelletal97-1}, which implies that the vast majority
of all CVs is expected to be dwarf novae. Based on the current
numbers, it appears that we know only a relatively small fraction of
the predicted short-period CV population. Another prediction made by
the population models is an accumulation of CVs near the minimum
orbital period \citep[e.g.][]{kolb+baraffe99-1}, which is not
observed. While this may signal a failure of the standard CV evolution
theory \citep[e.g.][]{kingetal02-1, barker+kolb03-1, andronovetal03-1},
it is likely that the known sample of CVs is incomplete and
biased. Dwarf novae are predominantly discovered because of their
outbursts \citep{gaensicke05-1}, and hence CVs with very long outburst
recurrence times or low-amplitude outbursts could be substantially
underrepresented in the known CV population.

As a measure to probe the completeness of the known CV sample, we have
initiated a search based on one property common to the majority of all
CVs: the presence of Balmer emission lines in their optical
spectra. We are selecting CV candidates from the Hamburg Quasar Survey
(HQS, \citealt{hagenetal95-1}), an objective prism Schmidt survey of
the northern hemisphere covering $13\,600\deg^2$ at high galactic
latitudes with a limiting magnitude $17.5\la B\la18.5$. The survey
resulted in $\simeq50$ new CVs, including a number of peculiar objects
\citep[e.g.][]{gaensickeetal00-2, rodriguez-giletal04-2,
rodriguez-giletal05-2, araujo-betancoretal05-1}; a general overview has
been given by \citet{gaensickeetal02-2} and more recently by
\citet{aungwerojwitetal05-1}. 

In this paper, we report the identification of five new dwarf novae in
the HQS: HS\,0417+7445, HS\,1016+3412, HS\,1340+1524, HS\,1857+7127,
and HS\,2214+2845 (HS\,0417, HS\,1016, HS\,1340, HS\,1857, and
HS\,2214, respectively, hereafter; Fig.\,\ref{f-fc} and
Table\,\ref{t-targets}). In Sect.\,\ref{s-observations} we provide
details about the observations and data reduction, in
Sect.\,\ref{hs0417}--\ref{hs2214} we describe the data analysis and
determine the orbital periods of the new dwarf novae. In
Sect.\,\ref{s-porbdn}, we compare the period distribution of the
dwarf novae found in the HQS to that of all known dwarf novae. In
Sect.\,\ref{s-spacedensity} we discuss the implications of our 
survey work on the space density of CVs.

\begin{table*}
\caption{\label{t-targets} Properties of the five new dwarf novae.}  \setlength{\tabcolsep}{1.1ex}
\begin{tabular}{lccccc}
\hline\noalign{\smallskip}
    & 
HS\,0417+7445 & 
HS\,1016+3412 & 
HS\,1340+1524 & 
HS\,1857+7127 &
HS\,2214+2845 \\
\hline\noalign{\smallskip}
Right ascension {(J2000)} & 
$04^\mathrm{h}23^\mathrm{m}32.8^\mathrm{s}$ &
$10^\mathrm{h}19^\mathrm{m}47.3^\mathrm{s}$ &
$13^\mathrm{h}43^\mathrm{m}23.2^\mathrm{s}$ &
$18^\mathrm{h}57^\mathrm{m}20.4^\mathrm{s}$ &
$22^\mathrm{h}16^\mathrm{m}31.2^\mathrm{s}$ \\
Declination {(J2000)} &
$+74\degr52\arcmin50.30\arcsec$ &
$+33\degr57\arcmin53.9\arcsec$ &
$+15\degr09\arcmin16.9\arcsec$ &
$+71\degr31\arcmin19.2\arcsec$ &
$+29\degr00\arcmin20.6\arcsec$ \\
Period (min) &
$\simeq105.1$/$\simeq109.9$ &
$114.3\pm2.7$ & 
$92.66\pm0.17$ &
$272.317\pm0.001$ &
$258.02\pm0.56$ \\
Magnitude range & 
$18.0-13.5$ & 
$18.6-15.4$: & 
$18.5-14.2$ & 
$17.2-13.9$ &
$16.5-12.3$ \\
\Ha\ EW [\AA] / FHWM [\AA] &
172 / 43 &
184 / 27 &
121 / 28 &
39  / 32 &
53  / 33 \\
\Hb\ EW [\AA] / FHWM [\AA] &
98 / 43 &
125 / 25 &
88 / 23 &
33 / 32 &
42 / 31 \\
\Hg\ EW [\AA] / FHWM [\AA] & 
73 / 38 &
85 / 24 &
59 / 22 & 
27 / 33 & 
30 / 32 \\
\Line{He}{I}{5876} EW [\AA] / FHWM [\AA] & 
40 / 43 &
52 / 31 & 
36 / 30 & 
7  / 34 & 
7  / 25 \\
\Line{He}{I}{6678} EW [\AA] / FHWM [\AA] &  
7  / 30 &
26 / 35 &
18 / 32 &
3  / 46 &
5  / 39 \\
RASS source (1RXS~J) &
042332.8+745300 &
101946.7+335811 &
134323.1+150916 &
185722.6+713126 &
221631.2+290025 \\
RASS count rate $\mathrm{(0.01\,cts\,s^{-1})}$ &
$6.0\pm1.3$ &
$2.5\pm1.0$ &
$7.3\pm1.7$ &
$3.4\pm4.5$ &
$9.9\pm1.5$ \\
Hardness ratio HR1 &
$1.00\pm0.09$ &
$1.00\pm0.91$ &
$0.18\pm0.23$ &
$1.00\pm0.04$ &
$0.92\pm0.06$ \\
Hardness ratio HR2 &
$0.21\pm0.20$ &
$-0.32\pm0.38$ &
$-0.05\pm0.27$ &
$0.69\pm0.10$ &
$0.10\pm0.14$ \\
\hline\noalign{\smallskip}
\end{tabular}

\parbox{\textwidth}{Notes. The coordinates are taken from the USNO-B
  catalogue \citep{monetetal03-1}; the ROSAT PSPC count rates and
  hardness ratios HR1 and HR2 have been obtained from the ROSAT All
  Sky Survey (RASS) Bright Source Catalogue \citep{vogesetal99-1} and
  from the RASS Faint Source Catalogue \citep{vogesetal00-1}; the
  \Ha--\Hg\ and \Line{He}{I}{5876, 6678} equivalent widths (EW) and
  full width at half maximum (FWHM) were measured from the Calar Alto
  average spectra (Fig.\,\ref{f-spectra}) using the
  \texttt{integrate/line} task in \texttt{MIDAS}; the outburst of
  HS\,1016 is uncertain (marked by a colon), as only one outburst
  was observed.}

\end{table*}

\section{Observations and Data Reduction\label{s-observations}}
\subsection{Spectroscopy}
\begin{figure*}
\centerline{\includegraphics[width=14cm]{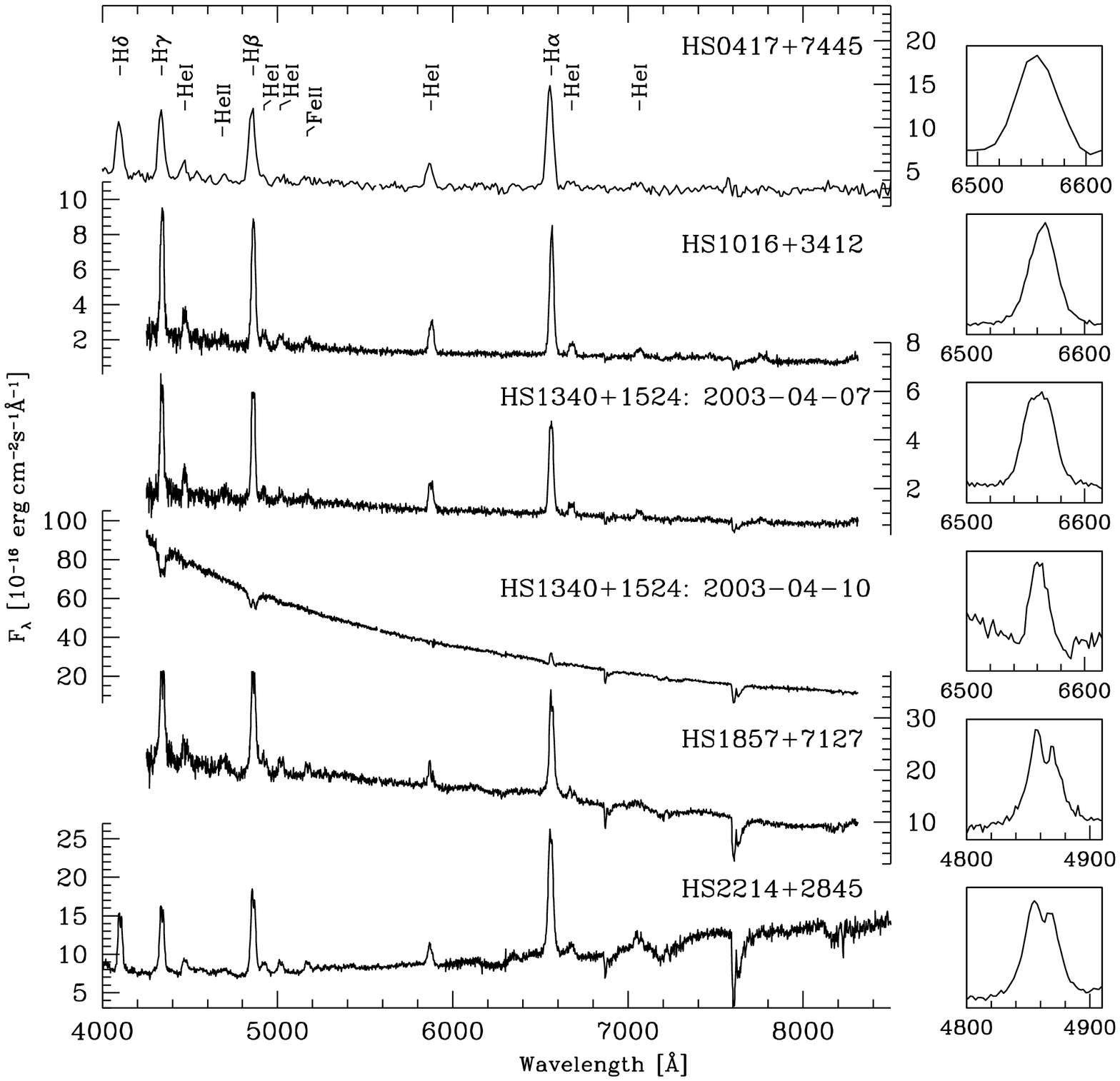}}
\caption{\label{f-spectra} Main panel: flux-calibrated CAFOS spectra
  of HS\,0417, HS\,1016, HS\,1340, HS\,1857, and HS\,2214. Fluxes are
  labelled alternatingly on the left and right side. HS\,1340 was
  observed in quiescence and outburst, respectively. Right panel:
  close-up plots of the \Ha\ and \Hb\ profiles.}
\end{figure*}

Identification spectra of HS\,0417, HS\,1016, HS\,1340,
HS\,1857, and HS\,2214 were obtained at Calar Alto Observatory
(Table\,\ref{t-obslog}).  The spectra of all five systems
(Fig.\,~\ref{f-spectra}) contain strong Balmer emission lines on a
blue continuum, together with weaker lines of \Idline{He}{I} that are
characteristic of CVs. \Line{He}{II}{4686} is very weak in all five
systems, suggesting that they are dwarf novae observed in quiescence.

Additional time-resolved spectroscopy of HS\,1016 (70 spectra),
HS\,1340 (78 spectra), HS\,1857 (41 spectra), and HS\,2214 (41
spectra) was obtained at the Calar Alto Observatory and  Roque de los
Muchachos Observatory (Table\,\ref{t-obslog}). The details of
instrument setup and data reduction are described below.

\paragraph{Calar Alto Observatory.}
Identification spectroscopy and time-resolved spectroscopy were
obtained with the Calar Alto Faint Object Spectrograph (CAFOS) at the
2.2\,m telescope throughout the period October 1996 to February 2005
(Table\,\ref{t-obslog}), with the exception of HS\,1857 which was
identified as a CV in 1990 using the B\&C spectrograph on the 2.2\,m
telescope. The identification spectra were obtained either with the
B-400 or the B-200 grating through a 1.5\,\arcsec slit on a
$2\mathrm{k}\times2\mathrm{k}$ pixel SITe CCD and were reduced with
the MIDAS quicklook context available at the Calar Alto.

The time-resolved follow-up observations of HS\,1016, HS\,1340,
HS\,1857, and HS\,2214 were obtained with the G-100 grating and a
1.2\,\arcsec slit, providing a spectral resolution of $\sim4.1$\,\AA\
(full width at half maximum, FWHM) over the wavelength range
$\lambda\lambda4240-8300$. Clouds and/or moderate to poor seeing
affected a substantial fraction of these observations.  HS\,2214 was
observed under photometric conditions using the B-100 grating along
with a 1.5\arcsec slit, providing a resolution of $\sim4$\,\AA\ (FWHM)
over the range 3500--6300\,\AA. Two additional red spectra of HS\,2214
were taken with the R-100 grating, covering the range 6000--9200\,\AA\
at a similar resolution. All follow-up spectroscopy was obtained in
600\,s exposures, interleaved with arc calibrations every
$\sim40$\,min to correct for instrument flexure. Flux standards were
observed at the beginning and end of the night~--~weather
permitting~--~ to correct for the instrumental response. The data
reduction (bias and flat-field correction, extraction, wavelength and
flux calibration) was carried out using the \texttt{Figaro} package
within Starlink and the programs \texttt{Pamela} and \texttt{Molly}
developed by T. Marsh. Special care was given to the wavelength
calibration by interpolating the dispersion relation for a given
target spectrum from the two adjacent arc exposures.

\paragraph{Observatorio del Roque de los Muchachos.} 
The Intermediate Dispersion Spectrograph (IDS) together with a
$2\mathrm{k}\times4\mathrm{k}$ pixel EEV10a detector was mounted at
the 2.5\,m Isaac Newton Telescope (INT) on La Palma to obtain
time-resolved spectroscopy of HS\,1016, HS\,1340, HS\,1857, and
HS\,2214 (Table\,\ref{t-obslog}). The R632V grating and a slit
width of 1.5\,\arcsec provided a spectral resolution of
$\sim2.3$\,\AA\ (FWHM) and a useful wavelength range of
$\sim4400-6800$\,\AA. The data reduction was carried out along the
same lines as described above for the Calar Alto data using
\texttt{IRAF}\footnote{\texttt{IRAF} is distributed by the National
Optical Astronomy Observatories.} and \texttt{Molly}. 

\begin{table*}
\setlength{\tabcolsep}{0.25ex}
\caption[]{Log of the observations\label{t-obslog}.}
\vspace*{-2.5ex}

\begin{minipage}[t]{9cm}
\begin{tabular}[t]{lcccccc}
\hline\noalign{\smallskip}
Date & UT &  Telescope & Filter/ & Exp. & Frames & Comp. \\    
     &    &            & Grism   & (s)  &        & star  \\    
\hline\noalign{\smallskip}
\multicolumn{7}{l}{\textbf{HS\,0417+7445}} \\
1996 Oct 05 & 02:52       & CA22 & B-400 & 600   & 1  & - \\
2000 Dec 21 & 22:41-02:37 & WD   & $B$   & 240   & 53 & C1 \\
2001 Jan 14 & 00:40-04:40 & WD   & $B$   & 240   & 54 & C1 \\
2003 Feb 27 & 20:35-23:47 & OLT  & $R$   & 300   & 37 & C2 \\
2004 Nov 12 & 17:40-02:03 & WS   & clear & 150   & 162 & C2 \\
2005 Jan 03 & 02:18-04:40 & INT  & $g^{\prime}$ & 80    & 71 & C2 \\
2005 Jan 04 & 23:07-02:27 & INT  & $g^{\prime}$ & 30-35 & 162 & C2 \\

\noalign{\smallskip}
\multicolumn {7}{l}{\textbf{HS\,1016+3412}} \\
2001 Apr 30 & 21:35       & CA22  & B-200 & 600   & 1   & - \\
2003 Apr 07 & 21:54       & CA22  & G-100 & 600   & 1   & - \\
2003 Apr 08 & 21:09-22:19 & CA22  & G-100 & 600   & 7   & - \\
2003 Apr 10 & 22:28       & CA22  & G-100 & 600   & 1   & - \\
2003 Apr 12 & 22:28-00:10 & CA22  & G-100 & 600   & 9   & - \\
2003 Apr 24 & 22:07-23:01 & INT   & R632V & 600   & 6   & - \\
2003 Apr 25 & 23:31-00:24 & INT   & R632V & 600   & 6   & - \\
2003 Apr 27 & 22:27-23:28 & CA22  & G-100 & 600   & 6   & - \\
2003 Apr 28 & 22:31-23:24 & INT   & R632V & 600   & 6   & - \\
2003 May 18 & 22:08-22:51 & INT   & R632V & 600   & 6   & - \\
2004 May 23 & 19:52-21:59 & KY    & clear & 45-90 & 104 & C3 \\
2004 May 24 & 20:09-21:47 & KY    & clear & 60    & 55  & C3 \\
2004 May 26 & 19:51-22:20 & KY    & clear & 45    & 157 & C3 \\
2004 May 27 & 19:58-23:10 & KY    & clear & 60-75 & 116 & C3 \\
2004 May 28 & 20:12-22:48 & KY    & clear & 60    & 101 & C3 \\
2004 May 29 & 22:52-23:57 & IAC80 & clear & 125   & 29  & C3 \\
2004 May 30 & 22:55-23:58 & IAC80 & clear & 120   & 28  & C3 \\
2004 May 31 & 21:18-22:48 & IAC80 & clear & 113   & 45  & C3 \\
2005 Feb 12 & 23:34-05:24 & CA22  & G-100  & 600  & 24  & C3 \\
\noalign{\smallskip}
\multicolumn{7}{l}{\textbf{HS\,1340+1524}}\\
2001 May 01 & 02:14       & CA22 & B-200 & 600 & 1   & - \\
2001 May 08 & 20:17-02:00 & AIP  & $R$   & 120 & 200 & C6 \\
2001 May 09 & 20:51-00:23 & AIP  & $R$   & 120 & 83  & C6 \\
2002 Jul 02 & 18:57-22:42 & KY   & $R$   & 120 & 86  & C4 \\
2002 Jul 04 & 20:02-22:17 & KY   & $R$   & 120 & 58  & C4 \\
2003 Apr 07 & 22:37-23:29 & CA22 & G-100 & 600 & 5   & - \\
2003 Apr 08 & 22:42-23:41 & CA22 & G-100 & 600 & 6   & - \\
2003 Apr 09 & 23:48-00:10 & CA22 & G-100 & 600 & 3   & - \\
2003 Apr 10 & 22:46-00:29 & CA22 & G-100 & 600 & 7   & - \\
2003 Apr 11 & 00:39-04:52 & CA22 & $V$   & 30  & 179 & C4 \\
2003 Apr 11 & 21:45-01:35 & CA22 & G-100 & 600 & 11  & - \\
2003 Apr 13 & 00:31-03:28 & CA22 & G-100 & 600 & 12  & - \\
2003 Apr 24 & 00:26-01:19 & INT  & R632V & 600 & 6   & - \\
2003 Apr 25 & 00:34-01:26 & INT  & R632V & 600 & 6   & - \\
2003 Apr 28 & 22:54-00:32 & CA22 & G-100 & 600 & 7   & - \\
2003 May 29 & 19:44-22:03 & KY   & $R$   & 90  & 69  & C4 \\
2003 May 30 & 19:17-01:12 & KY   & $R$   & 90  & 212 & C4 \\
2003 Jun 24 & 19:28-22:08 & KY   & $R$   & 90  & 101 & C4 \\
2003 Jun 28 & 19:21-22:10 & KY   & $R$   & 120 & 76  & C4 \\
2004 Mar 17 & 07:18-11:38 & FLWO & clear & 40  & 297 & C5 \\
\noalign{\smallskip}\hline
\end{tabular}
\end{minipage}
\hfill
\begin{minipage}[t]{9cm}
\begin{tabular}[t]{lcccccc}
\hline\noalign{\smallskip}
Date & UT &  Telescope & Filter/ & Exp. & Frames & Comp. \\    
     &    &            & Grism   & (s)  &        &  star \\    
\hline\noalign{\smallskip}
2004 Mar 18 & 07:23-12:00 & FLWO  & clear & 40    & 273 & C5 \\
2004 May 19 & 23:14-03:01 & IAC80 & clear & 60    & 193 & C6 \\
2004 May 20 & 22:43-00:58 & IAC80 & clear & 90    & 78  & C6 \\
2004 May 22 & 21:39-00:34 & KY    & clear & 30    & 262 & C4 \\
2004 Jun 10 & 20:01-21:54 & KY    & clear & 30    & 176 & C4 \\
2005 Feb 18 & 01:13-05:10 & CA22  & G-100 & 600   & 15  & - \\
2005 May 11 & 23:10-03:04 & IAC80 & clear & 45-60 & 203 & C6\\
2005 May 12 & 22:41-04:20 & IAC80 & clear & 60    & 213 & C6\\
2005 May 13 & 22:07-04:09 & IAC80 & clear & 60    & 270 & C6\\
\noalign{\smallskip}
\multicolumn{7}{l}{\textbf{HS\,1857+7127}} \\
1990 Jul 31 & 02:32       & CA22    & 120\,\AA/mm  & 3600 & 1 & - \\
2002 Apr 03 & 23:30-03:29 & AIP     & $R$   & 120  & 118  & C9 \\
2002 Apr 22 & 19:52-02:38 & AIP     & $R$   & 120  & 185  & C9 \\
2002 Sep 16 & 18:16-23:44 & KY      & $R$   & 5    & 1700 & C8 \\
2002 Sep 17 & 18:31-21:50 & KY      & $R$   & 5-10 & 683  & C8 \\
2002 Oct 24 & 18:05-04:27 & AIP     & $R$   & 60   & 447  & C9 \\
2003 Apr 08 & 02:44-03:44 & CA22    & G-100 & 600  & 6    & - \\
2003 Apr 09 & 02:10-04:47 & CA22    & G-100 & 600  & 10   & - \\
2003 Apr 13 & 02:25-04:50 & CA22    & G-100 & 600  & 9    & - \\
2003 Apr 22 & 19:45-03:04 & AIP     & $R$   & 60   & 386  & C9 \\
2003 Apr 25 & 04:01-05:03 & INT     & R632V & 600  & 7    & - \\
2003 Apr 27 & 05:32-05:43 & INT     & R632V & 600  & 2    & - \\
2003 Apr 29 & 04:18-04:30 & CA22    & G-100 & 600  & 2    & - \\
2003 Apr 29 & 04:46-05:28 & INT     & R632V & 600  & 5    & - \\
2003 Jul 08 & 03:18-05:19 & OGS     & clear & 20   & 279  & C7 \\
2003 Aug 17 & 22:42-23:08 & KY      & $V$   & 8    & 147  & C8 \\
2003 Aug 17 & 03:28       & HST/STIS & G140L & 800 & 1    & - \\
2004 May 03 & 00:21-05:46 & IAC80   & clear & 15   & 922  & C7 \\
2004 May 03 & 06:11-07:53 & FLWO    & clear & 10   & 487  & C7 \\
2004 May 04 & 08:25-11:15 & FLWO    & clear & 10   & 809  & C7 \\
\noalign{\smallskip}
\multicolumn{7}{l}{\textbf{HS\,2214+2845}}\\
2000 Sep 20 & 21:26       & CA22  & B-200 & 600 & 1   & - \\
2000 Sep 21 & 03:28-11:04 & BS    & $R$   & 100 & 245 & : \\
2000 Sep 24 & 02:44-10:53 & BS    & $R$   & 100 & 254 & : \\
2000 Sep 24 & 20:10-23:55 & CA22  & B-100 & 600 & 16  & - \\
2000 Sep 24 & 20:23-20:54 & CA22  & R-100 & 600 & 2   & - \\
2002 Aug 29 & 02:25-03:48 & INT   & R632V & 600 & 9   & - \\
2002 Sep 01 & 03:07-03:58 & INT   & R632V & 600 & 6   & - \\
2002 Sep 02 & 02:51-03:22 & INT   & R632V & 600 & 4   & - \\
2002 Sep 04 & 00:23-00:54 & INT   & R632V & 600 & 4   & - \\
2003 Jun 23 & 23:54-02:08 & KY    & $R$   & 90  & 85  & C11 \\
2003 Jun 25 & 00:06-02:11 & KY    & $R$   & 90  & 80  & C11 \\
2003 Jun 27 & 23:35-00:44 & KY    & $R$   & 30  & 113 & C11 \\
2003 Jun 28 & 22:49-00:35 & KY    & $R$   & 60  & 93  & C11 \\
2003 Jul 15 & 02:29-05:09 & OGS   & clear & 15  & 442 & C10 \\
2003 Sep 20 & 21:31-02:30 & IAC80 & clear & 10  & 871 & C10 \\
2003 Sep 21 & 19:53-02:52 & IAC80 & clear & 10  & 1187 & C10 \\
2003 Sep 22 & 20:14-03:10 & IAC80 & clear & 10  & 772  & C10 \\
2003 Sep 24 & 00:16-03:59 & IAC80 & clear & 10  & 689  & C10 \\
\noalign{\smallskip}\hline
\end{tabular}
\end{minipage}

\smallskip
Notes. CA22: 2.2\,m telescope at Calar Alto Observatory, using CAFOS
with a $2\mathrm{k}\times2\mathrm{k}$ SITe pixel CCD; WS: 0.8\,m
telescope at Wendelstein Observatory, using the MONICA CCD camera
\citep{roth90-1}; OLT: 1.2\,m Oskar L\"uhning Teleskop at Hamburg
Observatory, equipped with a $1\mathrm{k}\times1\mathrm{k}$ pixel SITe
CCD; INT: 2.5\,m Isaac Newton Telescope on Observatorio del Roque de
los Muchachos, equipped with the Wide Field Camera (WFC), an array of
four EEV $2\mathrm{k}\times4\mathrm{k}$ pixel CCDs; KY: 1.2\,m
telescope at Kryoneri Observatory, using a Photometrics SI-502
$516\times516$ pixel CCD camera; IAC80: 0.82\,m telescope at
Observatorio del Teide, equipped with Thomson
$1\mathrm{k}\times1\mathrm{k}$ pixel CCD camera; OGS: 1\,m Optical
Ground Station at Observatorio del Teide, equipped with Thomson
$1\mathrm{k}\times1\mathrm{k}$ pixel CCD camera; AIP: 0.7\,m telescope
of the Astrophysikalisches Institut Potsdam, using with
$1\mathrm{k}\times1\mathrm{k}$ pixel SITe CCD; FLWO: 1.2\,m telescope
at Fred Lawrence Whipple Observatory, equipped with the 4-Shooter CCD
camera, an array of four $2\mathrm{k}\times2\mathrm{k}$ pixel, only a
small part of the CCD\#3 was read out; BS: 0.41\,m at Braeside
Observatory telescope, using a SITe $512\times512$ pixel CCD camera;
the comparison stars used in instrumental magnitude extractions from
Braeside are unknown (marked by colons).
\end{table*}

%

\subsection{Photometry}
\begin{figure*}
\begin{minipage}[t]{\columnwidth}
\centerline{\includegraphics[angle=-90,width=\columnwidth]{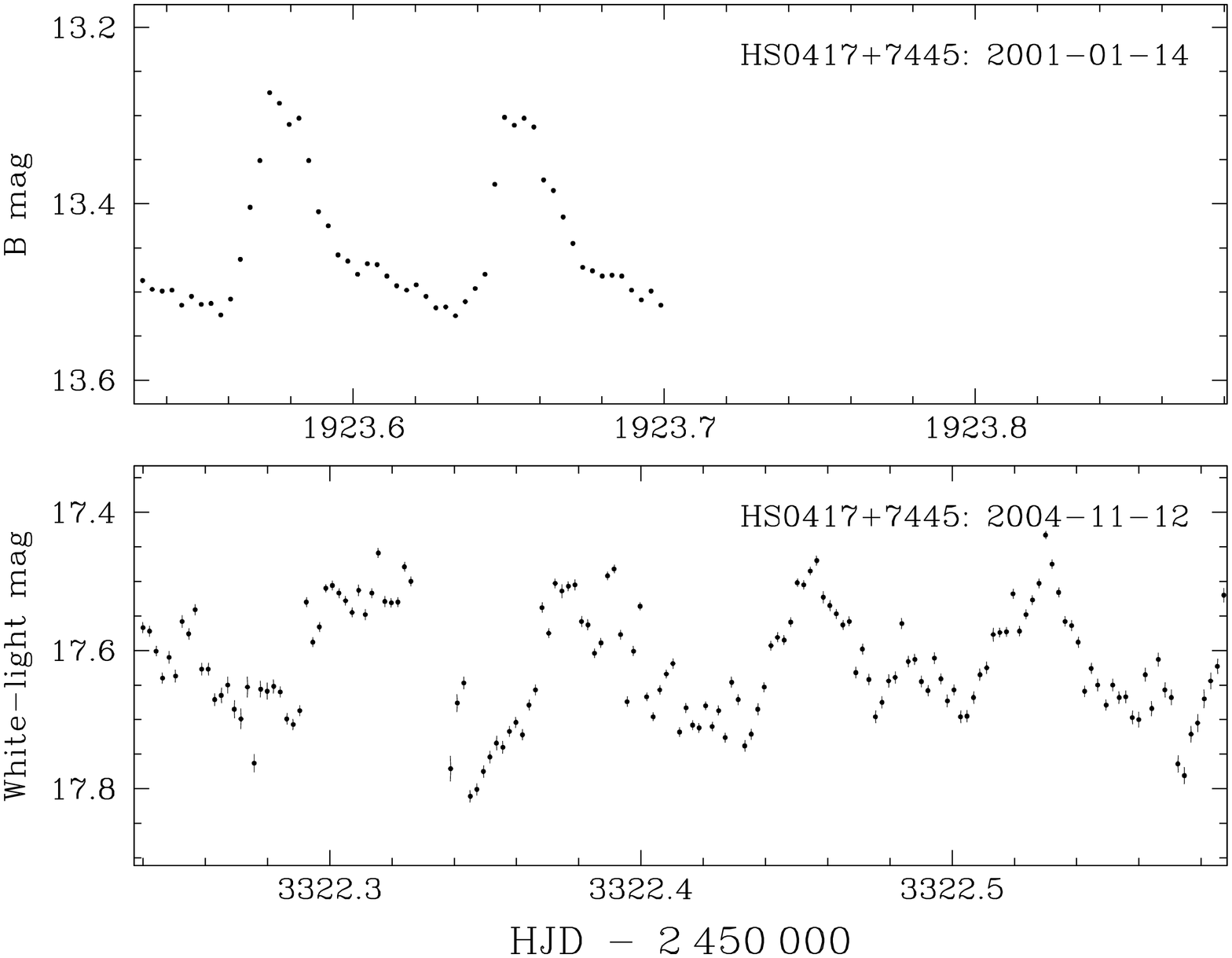}}
\caption [] {\label{f-lc_hs0417} Sample light curves of HS\,0417
obtained at the Wendelstein observatory. Top panel: $B$-band data
obtained during superoutburst on January 14, 2001. Bottom panel:
filterless data obtained during quiescence.}
\end{minipage}
\hfill
\begin{minipage}[t]{\columnwidth}
\centerline{\includegraphics[angle=-90,width=\columnwidth]{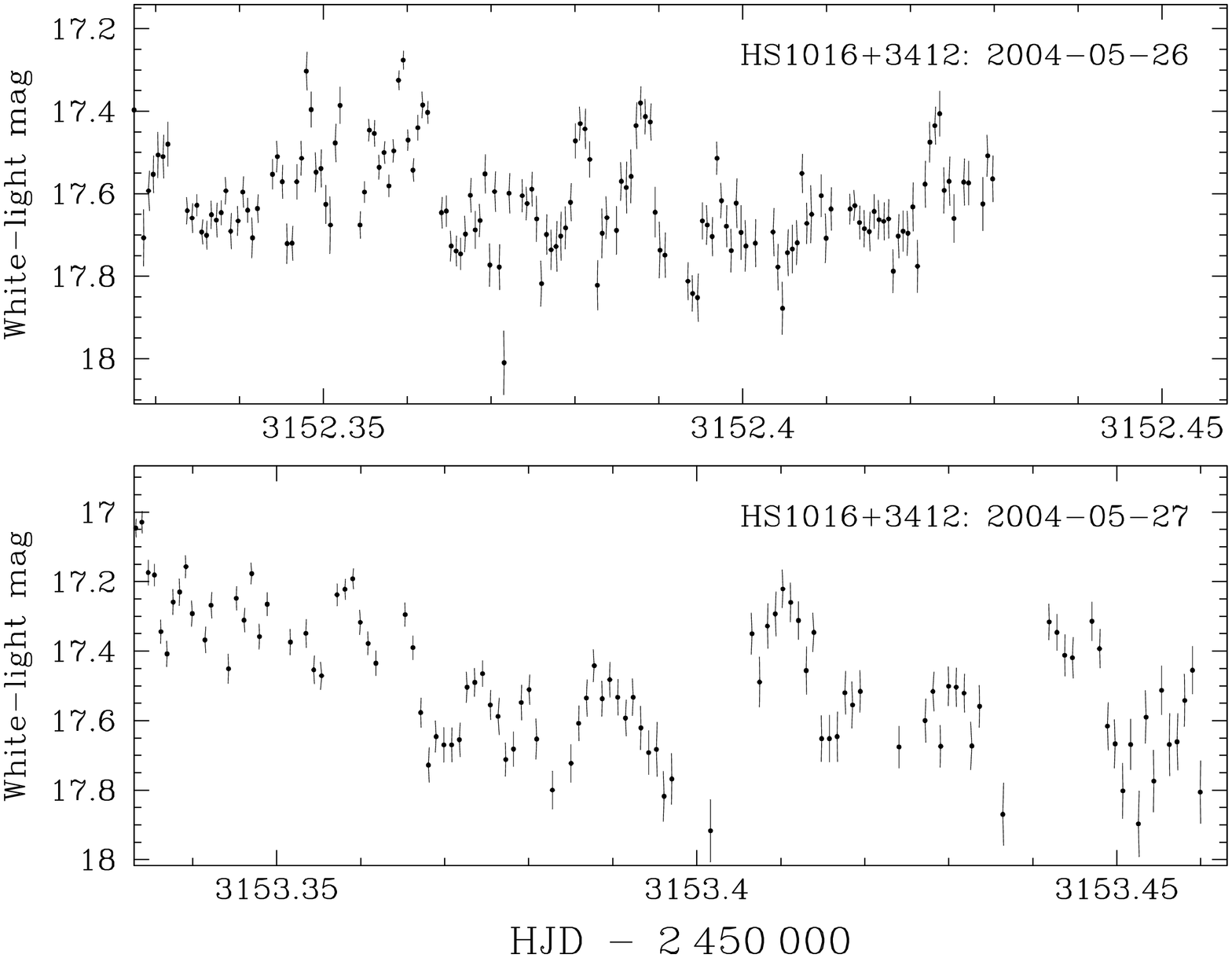}}
\caption [] {\label{f-lc_hs1016} Sample filterless light curves of
HS\,1016 obtained at the Kryoneri observatory.}
\end{minipage}

\medskip
\begin{minipage}[t]{\columnwidth}
\centerline{\includegraphics[angle=-90,width=\columnwidth]{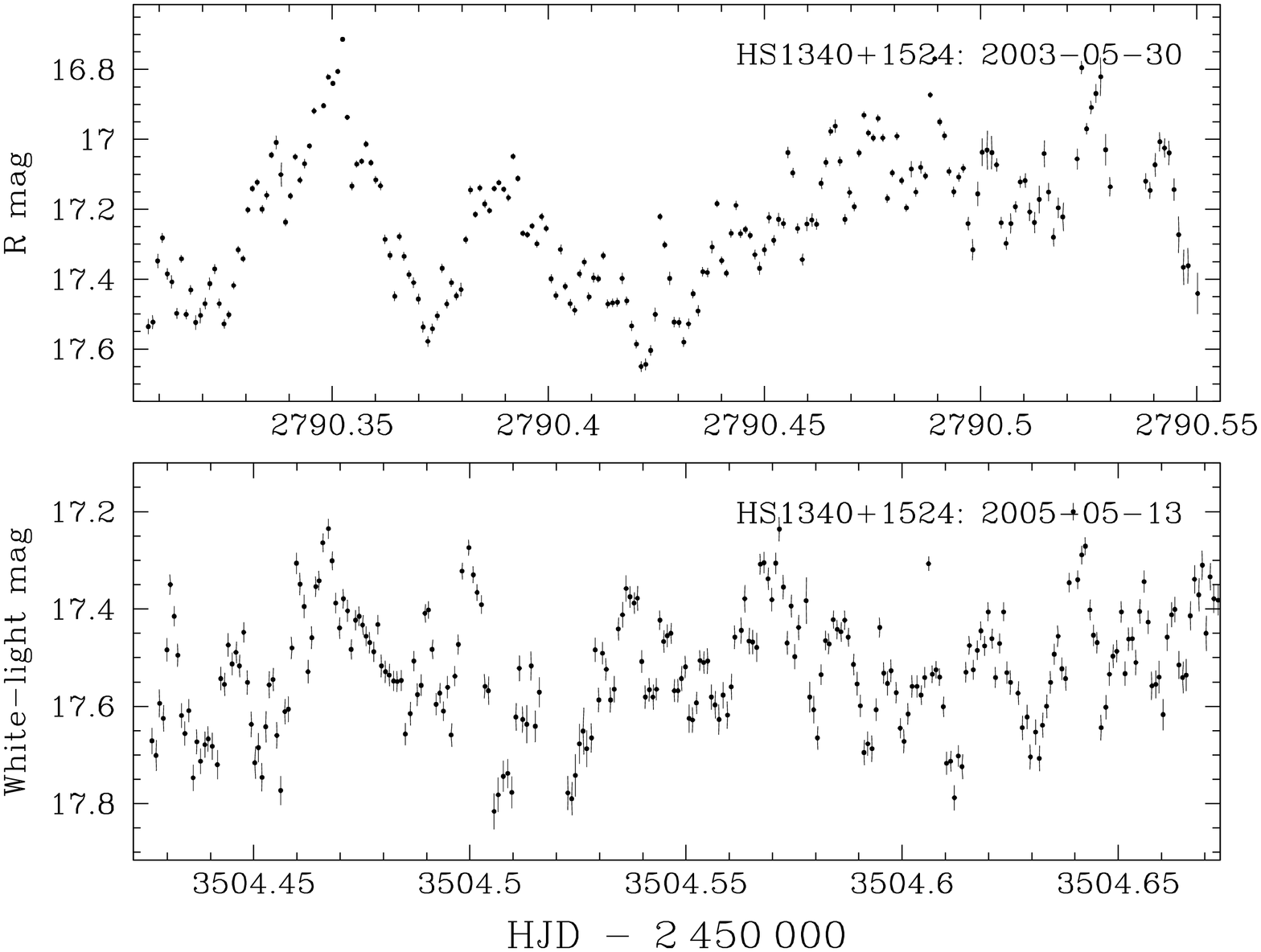}}
\caption [] {\label{f-lc_hs1340} Sample light curves of
HS\,1340. Top panel: $R$-band data obtained at the Kryoneri
observatory. Bottom panel: filterless data obtained at the IAC80
telescope.}
\end{minipage}
\hfill
\begin{minipage}[t]{\columnwidth}
\centerline{\includegraphics[angle=-90,width=\columnwidth]{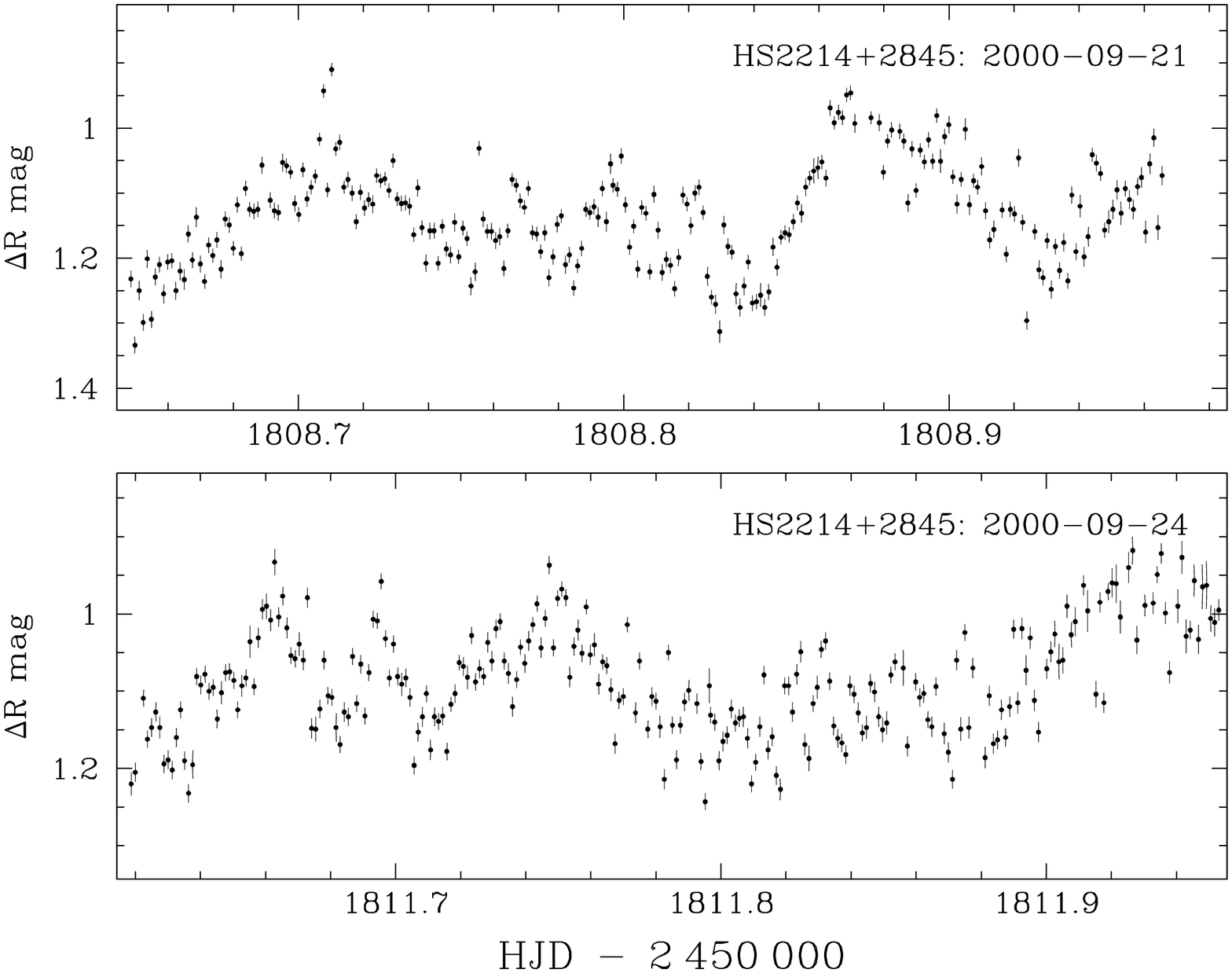}}
\caption [] {\label{f-lc_hs2214} Sample $R$-band light curves of
HS\,2214 obtained at the Braeside observatory.}
\end{minipage}
\end{figure*}

\begin{figure}
\centerline{\includegraphics[width=\columnwidth]{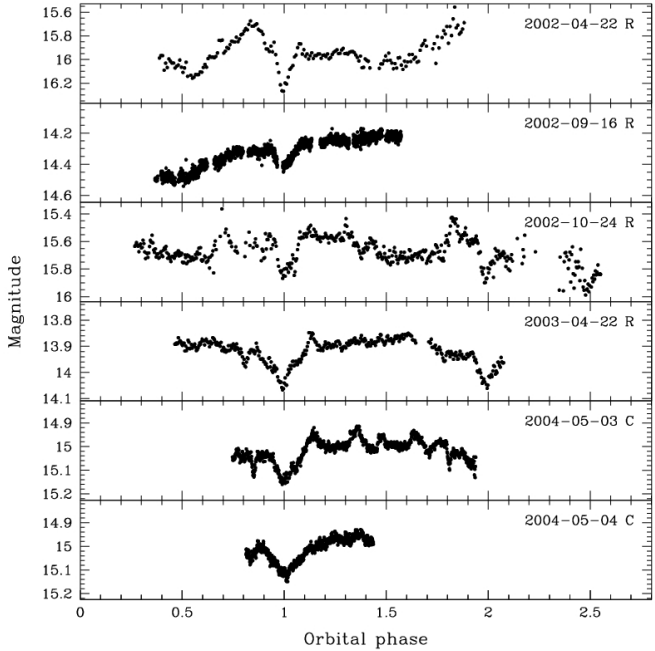}}
\caption [] {\label{f-fold_hs1857} Sample $R$-band and filterless
light curves of HS\,1857 folded over the ephemeris in
Eq.\,\ref{e-ephemeris}. See Sect.\,\ref{s-hs1857_phot} for details.}
\end{figure}

Throughout the period December 2000 to May 2005, we obtained
time-series differential CCD photometry of the five new dwarf novae
during a total of 54 nights using ten different telescopes. The
individual objects were observed for $\sim25$\,h (HS\,0417),
$\sim14$\,h (HS\,1016), $\sim68$\,h (HS\,1340), $\sim50$\,h
(HS\,1857), and $\sim48$\,h (HS\,2214). Sample light curves are shown
in Figs.\,~\ref{f-lc_hs0417} to \ref{f-fold_hs1857}. The
details of the observations and instruments used are given in
Table\,\ref{t-obslog}. The data obtained at Wendelstein, Calar Alto,
Kryoneri, INT, and OLT were reduced using the pipeline described by
\citet{gaensickeetal04-1}, which uses the \texttt{Sextractor}
\citep{bertin+arnouts96-1} to calculate aperture photometry for all
objects in the field of view. The AIP data were reduced entirely
within \texttt{MIDAS}. Bias and flat-field correction of the OGS,
IAC80, and FLWO images as well as the extraction of Point Spread
Function (PSF) magnitudes was done using \texttt{IRAF}. For the Braeside
data, the reduction was performed in a standard way using a
custom-made software suite. Finding charts of all five dwarf novae
are shown in Fig.\,\ref{f-fc}. The comparison stars used in the
reduction of our differential CCD photometry are listed in the
last column of Table\,\ref{t-obslog} (see Fig.\,\ref{f-fc} for
identifications), and their USNO $R$ and $B$ magnitudes are given in
Table\,\ref{t-compstars}.

Additional images of HS\,0417, HS\,1016,
HS\,1340, and HS\,2214 were taken intermittently during the period
May 2004 to April 2005 using the 0.37\,m robotic Rigel telescope of
the University of Iowa which is equipped with a
$1\mathrm{k}\times1\mathrm{k}$ pixel SITe-003 CCD camera. For all four
systems, filterless images with an exposure time of 25s were obtained.

\subsection{The new dwarf novae as X-ray sources}

All five dwarf novae identified on the basis of their emission line
spectra in the HQS are also X-ray sources in the ROSAT All Sky Survey (RASS):
HS\,0417, HS\,1340, and HS\,2214 are contained in the Bright Source
Catalogue \citep{vogesetal99-1}, HS\,1016 and HS\,1857 within the
Faint Source Catalogue \citep{vogesetal00-1}. The X-ray properties of
the new systems are summarised in Table\,\ref{t-targets}. All but
HS\,1340 are hard X-ray sources in the hardness ratio HR1, typical of
non- (or weakly-) magnetic CVs \citep{vanteeselingetal96-3}.

\begin{table}
\caption{\label{t-compstars} Comparison stars used for the
  differential CCD photometry of HS\,0417, HS\,1016, HS\,1340,
  HS\,1857, and HS\,2214, see Fig.\,\ref{f-fc}.}
\begin{tabular}{lccc}
\hline\noalign{\smallskip}
ID & USNO-A2.0 &  $R$ & $B$ \\
\hline\noalign{\smallskip}
C1 & 1575-02009718 & 13.3 & 13.3 \\
C2 & 1575-02008711 & 13.6 & 14.9 \\
C3 & 1200-06495553 & 14.3 & 15.0 \\
C4 & 1050-06991669 & 13.4 & 14.5 \\
C5 & 1050-06992410 & 14.4 & 16.4 \\
C6 & 1050-06992029 & 15.3 & 17.3 \\
C7 & 1575-04072972 & 12.2 & 13.5 \\
C8 & 1575-04073098 & 13.8 & 14.8 \\
C9 & 1575-04073991 & 13.8 & 14.9 \\
C10 & 1125-19198939 & 13.7 & 14.8 \\
C11 & 1125-19199670 & 15.0 & 15.6 \\
\hline\noalign{\smallskip}
\end{tabular}
\end{table}

\section{\label{hs0417}HS\,0417+7445}
Our identification spectrum of HS\,0417 obtained in October 1996
(Fig.\,\ref{f-spectra}, Table\,\ref{t-obslog}) is dominated by
low-excitation emission lines, typical of a dwarf nova. HS\,0417 is
contained in the ROSAT Bright Source Catalogue as 1RXS\,J042332+745300
\citep{vogesetal99-1}, and has been independently identified as a CV
by \citet{wuetal01-1}.  HS\,0417 displayed large-amplitude
variability on the HQS spectral plates, where it was detected at
$B\simeq18.0$ in June 1992 and at $B\simeq13.7$ in October 1995, supporting the
suggested dwarf nova nature of the object.

Throughout our photometric observations we have found the object near
a mean magnitude of $\simeq17.5$ (December 2000: $B\simeq17.9$,
February 2003: $R\simeq17.3$, November 2004: filterless $\simeq17.6$,
January 2005: $g^{\prime}\simeq17.5$), consistent with the USNO-A2.0
measurements, $R\simeq17.2$ and $B\simeq16.8$, except during January
2001, when the system was found in an outburst near $B\simeq13.5$. In
the quiescent state, the light curve of HS\,0417 is characterised by a
double-humped pattern with a period of $\sim100$\,min
(Fig.\ref{f-lc_hs0417}, bottom panel). The light curve obtained during
the January 2001 outburst (Fig.\,\ref{f-lc_hs0417}, top panel) reveals
superhumps that identify HS\,0417 as a SU\,UMa-type dwarf nova
and therefore this outburst as a superoutburst. An additional
outburst of HS\,0417 was caught on the rise in April 10, 2005 by one of
us (PS), and $\sim3$\,h, $V$-band data obtained by David Boyd on the
evening of April 11, 2005 showed the object already declining again at
a rate of $\sim0.85$ mag\,\id\ and no evidence of superhumps was
found. By April 18, the system reached again its quiescent magnitude
of $V\simeq17.5$.

\begin{figure} 
\centerline{\includegraphics[angle=-90,width=\columnwidth]
{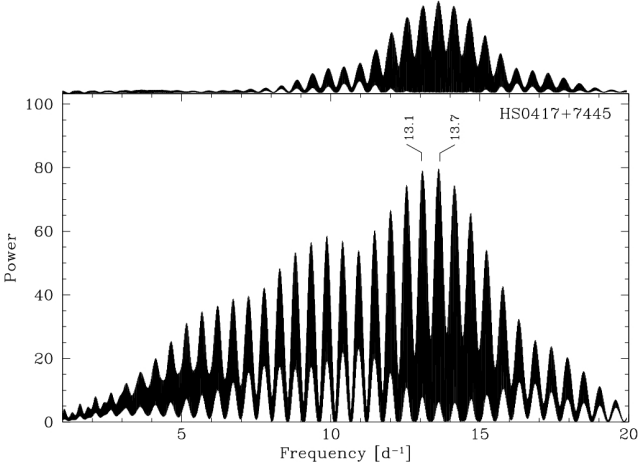}}
\caption [] {\label{f-scargle_hs0417} Main panel: the Scargle
periodogram of HS\,0417 during quiescence computed from all photometric data
except February 27, 2003. Top panel: the window function shifted to
13.7\,\id.}
\end{figure}

In order to measure the orbital period of the system, a
\citet{scargle82-1} periodogram was computed within the
\texttt{MIDAS/TSA} context from all quiescent data except the February
2003 observations which were of too poor a quality. The periodogram
(Fig.\,\ref{f-scargle_hs0417}) contains a fairly broad sequence of
aliases spaced by 1\,\id\ with the strongest signal at $13.7$\,\id\
and a nearly equally strong signal at $13.1$\,\id. The high-frequency
range of the periodogram of HS\,0417 is nicely reproduced by the
window function (shifted to 13.7\,\id\ in the top panel of
Fig.\,\ref{f-scargle_hs0417}), but excess power is present at
frequencies below 10\,\id, most likely associated with the short
length of the observing runs. Sine-fits to the data result in the
periods corresponding to the two highest peaks in the periodogram,
$P\simeq105.1$\,min and $P\simeq109.9$\,min, respectively.  We
interpreted these values as possible orbital periods of HS\,0417.

The Scargle periodogram computed from the superoutburst data obtained
on January 14, 2001 (Fig.\,\ref{f-lc_hs0417}, top panel) provides a
broad signal with a peak at $\simeq13.3$\,\id, or
$P\simeq108.3$\,min. The light curve folded over this period shows,
however, a significant offset between the two observed superhump maxima. A
periodogram computed using \citeauthor{schwarzenberg-czerny96-1}'s
(\citeyear{schwarzenberg-czerny96-1}) analysis-of-variance (AOV)
method using orthogonal polynomial fits to the data (implemented as
\texttt{ORT/TSA} in \texttt{MIDAS}) results in a much narrower peak
compared to the Scargle analysis, centred at $12.95$\,\id\
($P\simeq111.2$\,min). This period provides a clean folded
light curve. This improvement in the period analysis underlines the
fact that AOV-type methods provide better sensitivity for strongly
non-sinusoidal signals (such as superhumps) compared to
Fourier-transform based methods.

The analysis of our photometric data left us with two candidate
orbital periods, $\Porb=105.1$\,min or $\Porb=109.9$\,min, and two
candidate superhump periods, $\Psh=108.3$\,min or $\Psh=111.2$\,min.
Table\,\ref{t-epsilon} lists the fractional superhump excess,
$\epsilon=(\Psh-\Porb)/\Porb$ calculated from all possible
combinations of the candidate periods. We consider cases (2) and (3)
as very unlikely, as no dwarf nova with $\epsilon>5\%$ is found below
the period gap and no short-period dwarf nova with a negative
superhump excess is known \citep[e.g.][]{nogamietal00-1,
pattersonetal03-1, rodriguez-giletal05-1}. In fact, most dwarf novae
with $\Porb\sim100-115$\,min have $\epsilon\sim3-4\%$
\citep{pattersonetal05-3}, which would make case (1) look most
likely. However, based on our data, we prefer case (4) as
$\Psh=111.2$\,min gave the cleanest folded superhump light curve. In
this case, HS\,0417 would have a rather low value of $\epsilon$,
similar only to KV\,And ($\Porb=105.49(30)$\,min) which has
$\epsilon=0.0145$ \citep{pattersonetal03-1}. An unambiguous
determination of both $\Porb$ and $\Psh$ would be important, as
$\epsilon$ may be used to estimate the mass ratio of a CV
\citep{pattersonetal05-3}.

\begin{table} [t]
\caption[]{The fractional superhump excess of HS\,0417 computed from
$\epsilon=(\Psh-\Porb)/\Porb$\label{t-epsilon}.}
\begin{tabular}{cccc}
\hline\noalign{\smallskip} 
Case & \Porb\,(min) & \Psh\,(min) & $\epsilon$\\
\hline\noalign{\smallskip}
1 & 105.1 & 108.3 & 0.030\\
2 & 105.1 & 111.2 & 0.058\\
3 & 109.9 & 108.3 & -0.015\\
4 & 109.9 & 111.2 & 0.012\\
\noalign{\smallskip}\hline
\end{tabular}
\end{table}

\section{\label{hs1016}HS\,1016+3412}
The CAFOS (Fig.\,\ref{f-spectra}) and INT average spectra of HS\,1016
are similar to that of HS\,0417, with strong Balmer emission lines
together with weaker \Idline{He}{I} and \Idline{Fe}{II} lines and
practically absent \Line{He}{II}{4686}. Our photometric time-series
(Table\,\ref{t-obslog}) found the system consistently at a magnitude
of $\simeq17.5$. The system was found fainter, $V\simeq18.6$, in the April
2003 CAFOS acquisition images. The only known outburst of HS\,1016
was detected using the Rigel telescope on November 2, 2004, where an
unfiltered magnitude of 15.4 was recorded. The next image obtained on
November 11 showed the system again at its quiescent magnitude of
$\simeq17.5$.

The single-peaked profile found in the emission lines suggests a
relatively low orbital inclination. No spectral contribution
from the secondary star is detected in the red part of the
spectrum. The equivalent widths (EWs) from the CAFOS and INT average
spectra do not show any noticeable variation in each epoch throughout
our run. Table\,\ref{t-targets} lists FWHM and EW parameters of the
CAFOS average spectrum measured from Gaussian fits.

\begin{figure*}
\begin{minipage}[t]{\columnwidth}
\centerline{\includegraphics[width=\columnwidth]{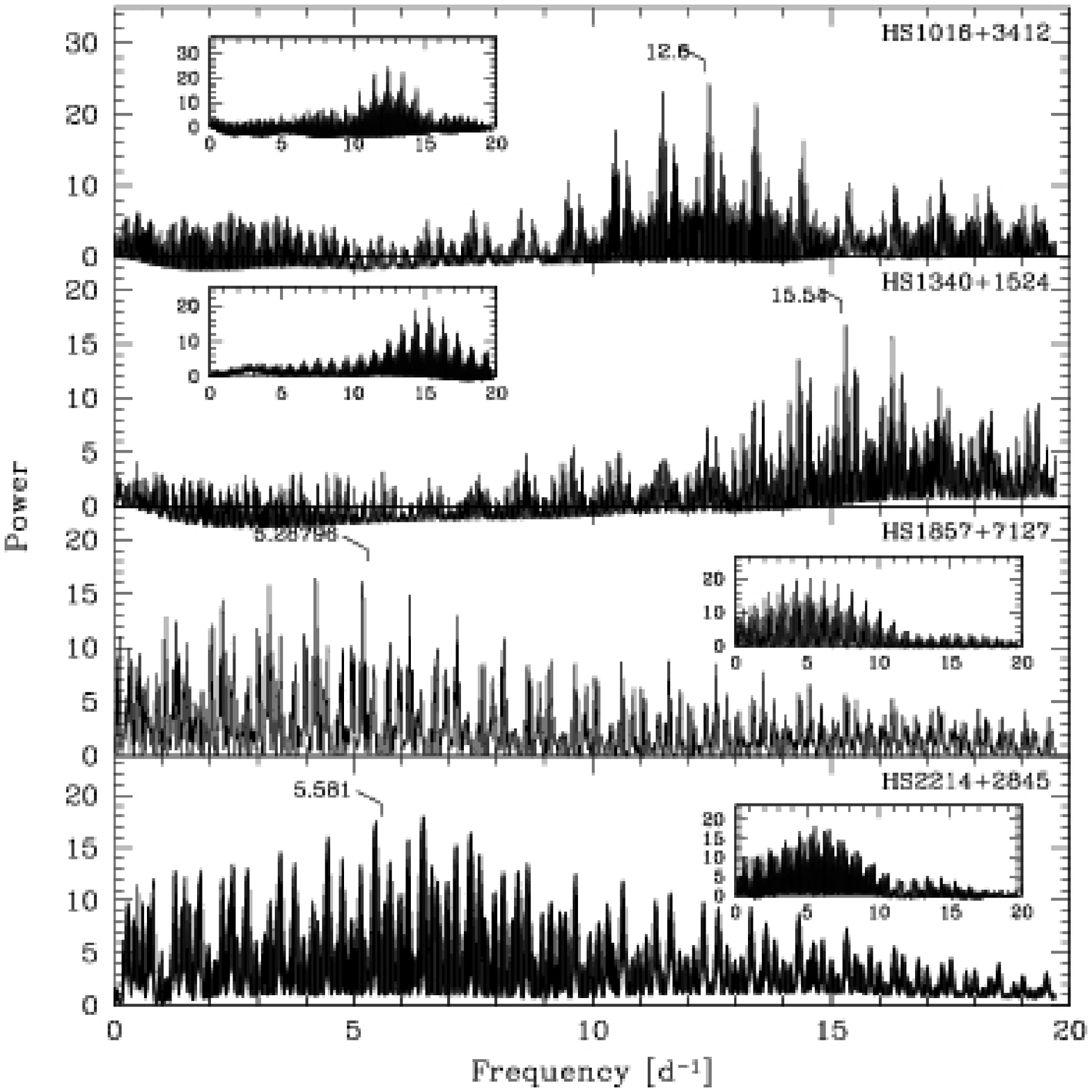}}
\caption [] {\label{f-scargle} The Scargle periodogram of the radial
velocities of HS\,1016, HS\,1340, HS\,1857, and HS\,2214. The
periodograms constructed from faked sets of data at the corresponding
orbital frequency are shown in small windows.}
\end{minipage}
\hfill
\begin{minipage}[t]{\columnwidth}
\centerline{\includegraphics[width=\columnwidth]{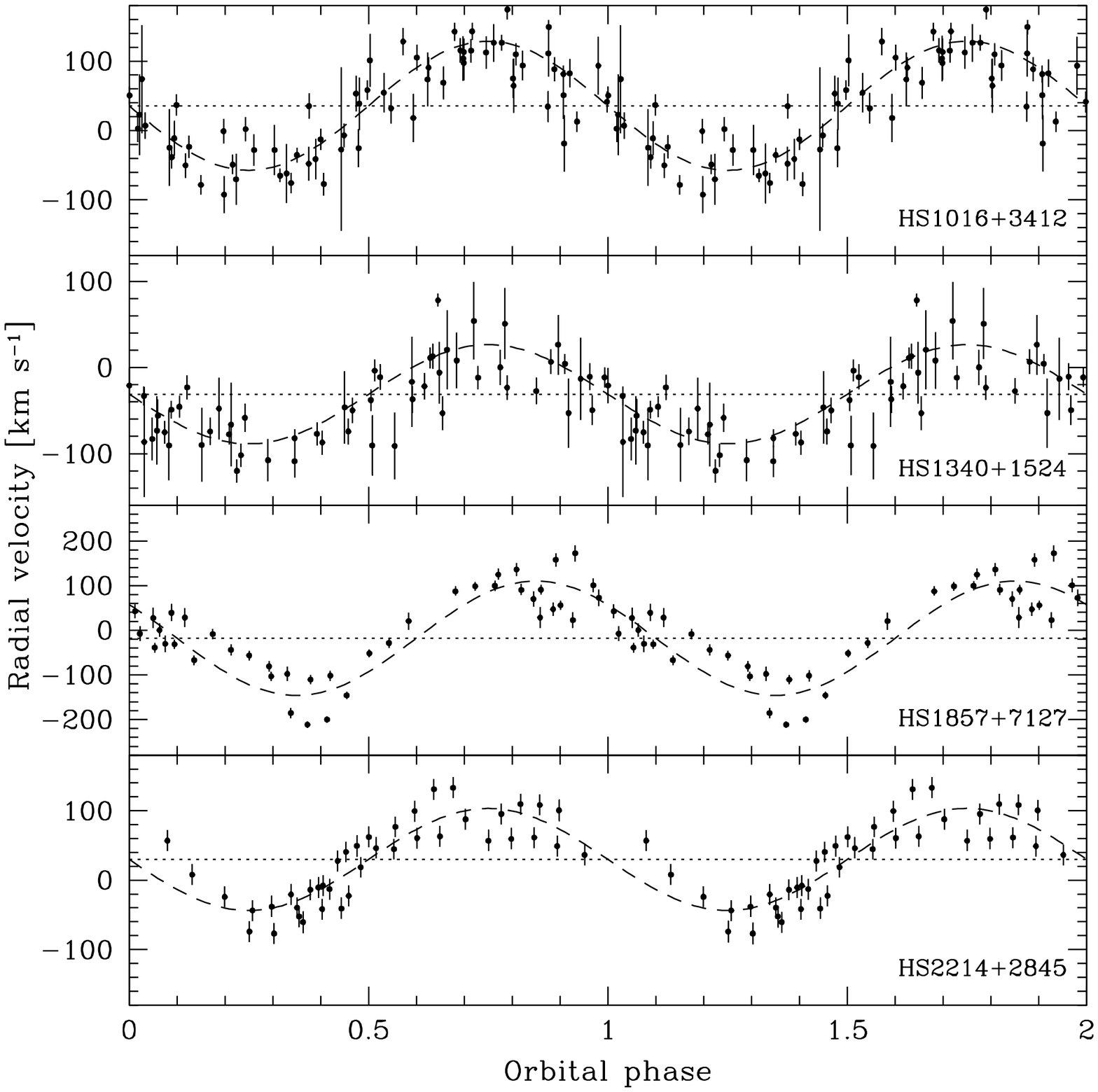}}
\caption [] {\label{f-rvfolded} \Ha\ radial velocities of HS\,1016,
    HS\,1340, and HS\,1857 and \Ha\ plus \Hb\ radial velocities of
    HS\,2214. The radial velocities of HS\,1016 and HS\,1340 are
    folded on their spectroscopic periods of 114.3\,min and
    92.66\,min, respectively and those of HS\,1857 and HS\,2214 folded
    on their photometric periods of 272.317\,min and 258.02\,min,
    respectively. The data on HS\,1857 were folded using the eclipse
    ephemeris given in Eq.\,\ref{e-ephemeris} while the other three
    systems were folded defining phase zero as the time of red-to-blue
    crossing of the radial velocities.}
\end{minipage}
\end{figure*}

In order to determine the orbital period of HS\,1016, we measured
the radial velocity variation of \Ha, the strongest emission line,
from the CAFOS and IDS spectra. We first rebinned the individual
spectra to a uniform velocity centred on \Ha, followed by normalising
the slope of the continuum. We then measured the \Ha\ radial velocity
variation using the double Gaussian method of
\citet{schneider+young80-2} with a separation of 1000\,\kms\ and an
FWHM of 200\,\kms. A Scargle periodogram calculated from the \Ha\
radial velocity variation contains a set of narrow aliases spaced by
1\,\id, with the strongest signal found at $f\simeq12.6$\,\id\
(Fig.\,\ref{f-scargle}, top panel). We tested the significance of this
signal by creating a faked set of radial velocities computed from a
sine function with a frequency of $12.6$\,\id, and randomly offset
from the computed sine wave using the observed errors. The periodogram
of the faked data set is plotted in a small window of the top panel in
Fig.\,\ref{f-scargle} which reproduces well the alias structure of the
periodogram calculated from the observation. A sine-fit to the folded
radial velocities refined the period to $114.3\pm2.7$\,min, which we
interpreted as the orbital period of
HS\,1016. Figure\,\ref{f-rvfolded} (top panel) shows a sine-fit to 
the phase-folded radial velocity curve; the fit parameters are
reported in Table\,\ref{t-rvfits}.

The light curves of HS\,1016 display  short-time scale flickering
with an amplitude of 
$\sim0.2-0.3$\,mag (Fig.\,\ref{f-lc_hs1016}). A  Scargle periodogram
computed from the entire photometry as well as from individual subsets
did not reveal any significant signal.

\begin{table*} [t]
\caption[]{Sine fits to the \Ha\ radial velocities of
HS\,1016, HS,1340, and HS\,1857. For HS\,2214 a combination of \Ha\
and \Hb\ radial velocities were fitted, as the September 2000 Calar
Alto spectra did not cover \Ha. For HS\,1857 and HS\,2214, the periods
were fixed to their values determined from the photometry. \label{t-rvfits}}
\begin{flushleft}
\begin{tabular}{ccccc}
\hline\noalign{\smallskip} Object & \T\ & Period
(days) & K (\kms) & $\gamma$ (\kms)\\ \hline\noalign{\smallskip}
HS\,1016+3412 & $2452737.4039\pm0.0012$ & $0.0794\pm0.0019$ &
$93.0\pm5.3$ & $35.6\pm3.9$ \\ 
HS\,1340+1524 & $2452737.4438\pm0.0018$ & $0.06435\pm0.00012$ & 
$57.5\pm6.2$  & $-31.1\pm4.1$ \\ 
HS\,1857+7127 & $2452368.53243\pm0.00098$ & $0.189109\pm0.000001$ & 
$128.0\pm9.6$ & $-17.9\pm7.4$ \\ 
HS\,2214+2845 & $2451812.3309\pm0.0028$ & $0.17918\pm0.00039$ & 
$73.5\pm6.5$  & $29.8\pm5.0$ \\
\noalign{\smallskip}\hline
\end{tabular}
\end{flushleft}
\end{table*}

\section{\label{hs1340}HS\,1340+1524}
The average spectrum of HS\,1340 during quiescence
(Fig.\,\ref{f-spectra}) is similar to that of HS\,0417 and HS\,1016,
showing strong single-peaked line profiles of Balmer emissions along
with the weaker lines of \Idline{He}{I} and \Idline{Fe}{II}. The
line parameters during quiescence are given in Table\,\ref{t-targets}.

\subsection{Long and short term variability}
\begin{figure} 
\centerline{\includegraphics[angle=-90,width=\columnwidth]{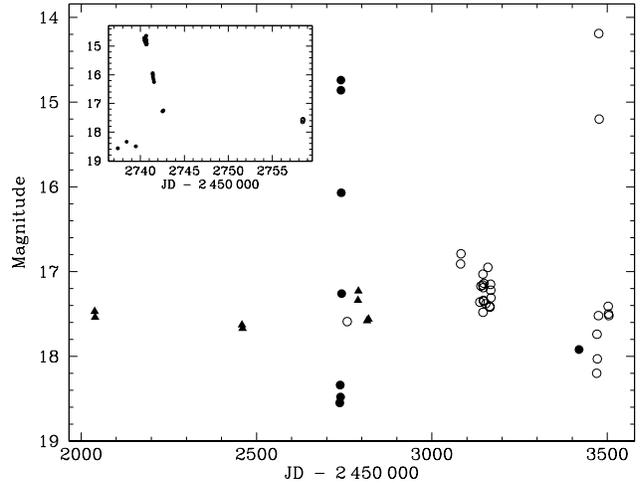}}
\caption [] {\label{f-hs1340_mag} Main window: the mean magnitudes of
HS1340 obtained from May 2001 to May 2005 in $R$-band (filled
triangles), $V$-band (filled circles), and white light (open circles). The
photometric error on the individual points is $<0.05$\,mag. An
additional systematic uncertainty arises from the combination of
different band passes. Considering the apparent magnitudes of HS\,1340
listed in the Sloan Digital Sky Survey which are $g=17.3$, $r=17.1$,
and $i=17.1$, the errors due to colour terms are likely to be within
$\pm0.1$\,mag.  Small window: close up of the April\,2003 CAFOS
run. The first three points show the deep faint state with a mean
magnitude of $V\simeq18.5$ before the outburst on April 10, 2003.}
\end{figure}

Throughout our time series photometry obtained at the AIP, IAC80,
FLWO, and Kryoneri, HS\,1340 was found at a mean magnitude in the
range $\sim17.7-16.8$ (see Fig.\,\ref{f-hs1340_mag}, main window).  A
first outburst of HS\,1340 was detected on the evening of April 10,
2003 during observations with the 2.2m telescope at the Calar Alto Observatory
(Table\,\ref{t-obslog}). The outburst peak magnitude was
$V\simeq14.7$ on the CAFOS acquisition image. The spectra obtained
immediately thereafter show weak emission at \Ha, with an equivalent
width of $\sim4.7$\,\AA, whereas \Hb\ and \Hg\ are in absorption with
narrow emission cores, which are typical of an optically thick
accretion disc.  Similar spectra were observed in the three HQS CVs
HS\,0139+0559, HS\,0229+8016, and HS\,0642+5049
\citep{aungwerojwitetal05-1}.  As the conditions during the night
deteriorated, we switched to time-series photometry, recording a
decline at $\sim0.05\,\mathrm{mag\,hr^{-1}}$. The CAFOS acquisition images
showed that HS\,1340 faded to $V\simeq16.1$ and $V\simeq17.3$ in the
two subsequent nights, April 11 and 12, 2003, respectively. A puzzling
fact is that acquisition images taken before the outburst on April 7,
8, and 9, 2003 showed HS\,1340 at $V\simeq18.5$, i.e. nearly one
magnitude fainter than the usual quiescent value (see
Fig.\,\ref{f-hs1340_mag}, small window). On April 28, a CAFOS
acquisition image showed the system again at a filterless magnitude of
17.6, consistent with the typical quiescent brightness. The duration
of the entire outburst was less than two days.

A second outburst reaching an unfiltered magnitude of $\sim14.2$ was
recorded on April 15, 2005 with the Rigel telescope, again, the
duration of the outburst was of the order of $2-3$ days. 

The light curves of HS\,1340 obtained during quiescence are
predominantly characterised by variability on time scales of
$\sim15-20$\,min with peak-to-peak amplitudes of $\sim0.4$\,mag
(Fig.\ref{f-lc_hs1340}, bottom panel). On some occasions, the light
curves shows hump-like structures which last for one to several hours,
superimposed by short-time scale flickering
(e.g. Fig.\ref{f-lc_hs1340}, top panel). Our period analysis of the
photometric data did not reveal any stable signal in the combined data.

In summary, HS\,1340 appears to have rather infrequent and short-lived
outbursts, and displays a substantial amount of short-term variability
as well as variability of its mean magnitude during quiescence.

\subsection{The orbital period}

The orbital period of HS\,1340 was determined using the spectroscopic
data taken in quiescence. The \Ha\ radial velocity variation was
measured in the same manner as in HS\,1016 with a separation of
1000\,\kms\ and an FWHM of 200\,\kms. Figure\,\ref{f-scargle} (second
panel) shows the Scargle periodogram.  The strongest signal is found
at $f=15.54\pm0.03$\,\id\ where the error is estimated from the FWHM
of the strongest peak in the periodogram, corresponding to an orbital
period of $\Porb=92.66\pm0.17$\,min. The \Ha\ radial velocity curve folded
over this period is shown in Fig.\,\ref{f-rvfolded} (second panel)
along with a sine-fit; the fit parameters are given in
Table\,~\ref{t-rvfits}.  The periodogram of a faked data set
constructed from this frequency agrees well with the entire observed
alias structure (insert in Fig.\,\ref{f-scargle}, second panel).

With the spectroscopic period being determined, we re-analysed the
time-series photometry of HS\,1340, and found no significant signal in
the range of the orbital frequency when we combined all quiescent
data. However, a weak signal at a frequency of $\sim15.5$\,\id\ and
its one-day aliases were detected intermittently on some occasions,
e.g. in the 2003 Kryoneri data and the 2004 FLWO observations.

\section{\label{hs1857}HS\,1857+7127}
The CAFOS average spectrum of HS\,1857 (Fig.\,\ref{f-spectra}) is
similar to the spectra of HS\,0417, HS\,1016, and HS\,1340, presenting
a blue slope superimposed by Balmer and \Idline{He}{I} emission
lines. Slight flux depressions are observed near 6200\,\AA\ and
7200\,\AA, which might be associated with the TiO bands of an M-type
donor, however, the quality of the data is insufficient to
unambiguously establish the detection of the secondary star. The
Balmer emission line profiles are double-peaked, with a peak-to-peak
separation of $\sim800$\,\kms, suggesting a moderate to high
inclination of the system. A high orbital inclination of HS\,1857 was
confirmed by the detection of eclipses in the light curves of the
system (Fig.\,\ref{f-fold_hs1857}).

\subsection{Long term variability}
\begin{figure}
\centerline{\includegraphics[angle=-90,width=\columnwidth]{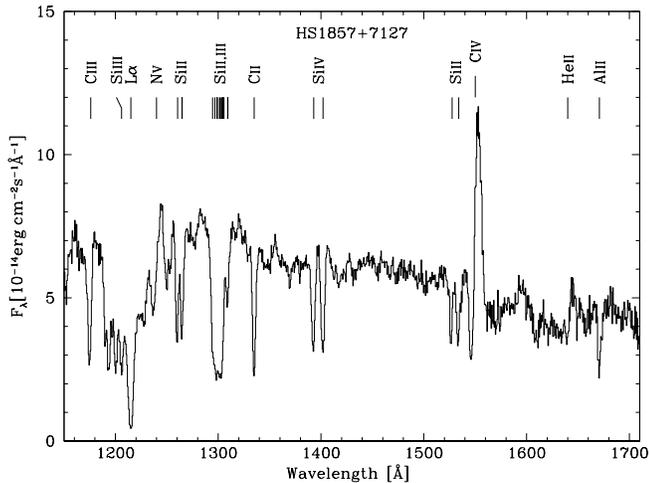}}
\caption [] {\label{f-hs1857_stis} The HST/STIS spectrum of HS\,1857
  taken on August 17, 2003 during an outburst.}
\end{figure}

Throughout our photometric observing runs, HS\,1857 was found to vary
over a relatively large range between $17.2-13.9$\,mag in average
brightness, suggesting a frequent outburst activity
(Fig.\,\ref{f-fold_hs1857}). Combined with the long orbital period
(see below), it appears likely that HS\,1857 is a Z\,Cam-type dwarf
nova. The INT spectra obtained in April 2003 showed the system with a
broad absorption trough around \Hb, with a weak
($\mathrm{EW}\sim3.5$\,\AA) single-peaked emission core, typical of
dwarf novae during outburst \citep{hessmanetal84-1}. As we did not
obtain a spectrophotometric flux standard on that occasion, and
have no simultaneous photometric data, the magnitude of that outburst
could not be determined. An additional outburst spectrum was obtained
in the ultraviolet using HST/STIS on August 17, 2003, showing a range
of low and high ionisation lines of C, N, Si, and Al in absorption, as
well as a P-Cygni profile in \Line{C}{IV}{1550}
(Fig.\,\ref{f-hs1857_stis}). We derived an $R$-band equivalent
magnitude of 14.1 from the STIS acquisition image taken before the
ultraviolet spectroscopy (see \citealt{araujo-betancoretal05-2} for
details on the processing of STIS acquisition images), and
ground-based photometry obtained at Kryoneri a few hours after the
STIS observations found HS\,1857 at $V=13.9$. The STIS spectrum
resembles qualitatively the ultraviolet spectrum of Z\,Cam obtained
during an outburst \citep{kniggeetal97-1}. The P-Cygni profile
provides evidence for the presence of a wind outflow during the
outburst.

\subsection{Eclipse ephemeris}
\label{s-hs1857_phot}
\begin{figure}
\centerline{\includegraphics[angle=-90,width=\columnwidth]{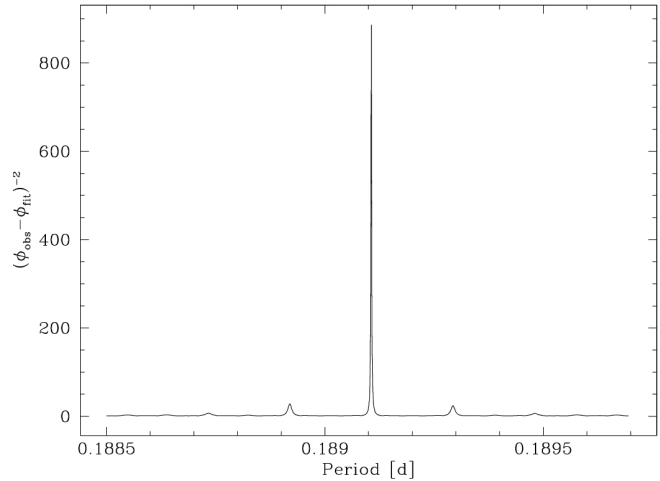}}
\caption [] {\label{f-periodogram_hs1857} The periodogram of HS\,1857
computed from nine eclipses obtained during the 2002 to 2004 runs.}
\end{figure}

We obtained light curves of HS\,1857 throughout the period 2002 and
2004 and covered nine eclipses. We measured the times of the eclipse
minima, and determined the cycle count by fitting
$(\phi_0^\mathrm{\mathrm{fit}}-\phi_0^{\mathrm{observed}})^{-2}$ ,
leaving the period as a free parameter
(Fig.\,\ref{f-periodogram_hs1857}). The following ephemeris was
derived as:
\begin{equation}
\label{e-ephemeris}
\phi_0=\mathrm{HJD}\,2452368.53243(98)+ 0.189109(1)\times E
\end{equation}
where $\phi_0$ is defined as the phase of mid-eclipse. The errors
(given in brackets) of the zero phase and period were determined from
a least-squares fit to the observed eclipse times versus the cycle
count number. We conclude that the orbital period of HS\,1857 is
$\Porb=272.317(1)$\,min. 

\begin{table} [t]
  \caption[]{The times of eclipse minima of HS\,1857 obtained
during the 2002 to 2004 runs \label{t-eclipse_minima}.}
\begin{tabular}{cc}
\hline\noalign{\smallskip} 
Date & Eclipse minima (HJD) \\
\hline\noalign{\smallskip}
2002 Apr 03 & 2452368.53446 \\
2002 Apr 22 & 2452387.44260 \\
2002 Sep 16 & 2452534.38100 \\
2002 Oct 24 & 2452572.39500, 2452572.57893 \\
2003 Apr 22 & 2452752.42275, 2452752.61229 \\
2004 May 03 & 2453128.56222 \\
2004 May 04 & 2453129.88816 \\
\noalign{\smallskip}\hline
\end{tabular}
\end{table}

The overall shape of the light curves and that of the eclipse profiles
show a large degree of variability (Fig.\,\ref{f-fold_hs1857}). On
April 22, 2002, the light curve shows an orbital modulation with a
bright hump preceding the eclipse, typically observed in
quiescent eclipsing dwarf novae \citep[e.g.][]{zhangetal87-1},
produced by the bright spot.  A shallow ($\sim0.4$\,mag) eclipse is
recorded, implying a partial eclipse of the accretion disc in the
system. On September 16, 2002, the system was apparently caught on the
rise to an outburst, with the eclipse depth reduced to
$\sim0.2$\,mag. During several intermediate and bright states the
signature of the bright spot disappeared, and was replaced by a broad
orbital modulation with maximum light near phase 0.5, superimposed by
short time scale flickering (Fig.\,\ref{f-fold_hs1857}, bottom four
panels). On May 3, 2004, a narrow dip ($\Delta\phi\simeq0.05$) centred
at $\phi\simeq0.8$ preceds the eclipse during both observed cycles. A
similar feature, though of lower depth, has been observed on April 22,
2003. 

\subsection{Radial velocities}
As for HS\,1016 and HS\,1340, we measured the \Ha\ radial
velocities of HS\,1857 using the double-Gaussian method with a
separation of 1500\,\kms\ and an FWHM of 400\,\kms. The Scargle
periodogram computed from these data contains a set of narrow peaks at
3.3\,\id, 4.3\,\id, and 5.3\,\id, consistent with the photometric
frequency, $f\simeq5.29$\,\id, computed from the eclipse ephemeris in the
previous section, and its one-day aliases (Fig.\,\ref{f-scargle},
third panel). A periodogram calculated from a faked data set assuming the
photometric frequency of $5.28796$\,\id, is shown in the insert in
Fig.\,\ref{f-scargle} (third panel). We folded the \Ha\ radial
velocity curve using the eclipse ephemeris given in
Eq.\,\ref{e-ephemeris}, resulting in a quasi-sinusoidal modulation
with an amplitude of $128.0\pm9.6$\,\kms\ and
$\gamma=-17.9\pm7.4$\,\kms, as determined from a sine-fit
(Table\,\ref{t-rvfits}). The red-to-blue crossing of the \Ha\
radial velocities occurs at the photometric phase $\sim0.1$
(Fig.\,\ref{f-rvfolded}, third panel).  Such a shift is not too much
of a surprise, as our radial velocity measurements were extracted from
spectra sampling different brightness (outburst) states of HS\,1857,
covering less than one orbital cycle in all cases, and are not
expected to represent a uniform and symmetrical emission from the
accretion disc.

\section{\label{hs2214}HS\,2214+2845}
The CAFOS average spectrum of HS\,2214 (Fig.\,~\ref{f-spectra}) is
characterised by a fairly red continuum superimposed by strong Balmer
emission lines and weaker emission lines of \Idline{He}{I},
\Idline{He}{II}, and \Idline{Fe}{II}. The Balmer line
profiles are double-peaked with a peak-to-peak separation of
$\sim800-1000$\,\kms, suggesting an origin in an accretion disc
\citep{horne+marsh86-1}.  TiO absorption bands are present in the red
part of the spectrum, revealing the late-type secondary star. 

During our spectroscopic and photometric follow-up studies, HS\,2214
was consistently found at $16.5-15.5$\,mag, with night-to-night
variations of the mean magnitude of $0.2-0.4$\,mag. The dwarf nova
nature of HS\,2214 was confirmed through the visual monitoring by one
of us (PS), which led to the detection of outbursts on December 10,
2004, and on May 2, July 14, September 22, 2005, and April 23,
2006. The mean cycle length appears to be $\sim71$\,d, and the maximum
brightness recorded during outburst was $V\simeq12.3$.

\subsection{The spectral type of the secondary and distance} 
Overall, the spectrum of HS\,2214 resembles that of the U\,Gem-type
dwarf novae, e.g. CZ\,Ori \citep{ringwaldetal94-1}, PG\,0935+075
\citep{thorstensen+taylor01-1}, and U\,Gem itself
\citep{wade79-1,staufferetal79-1}, which have secondary stars with
spectral types in the range M2--M4, and orbital periods in the range
255--315\,min. 

In order to determine the spectral type of the secondary star in
HS\,2214, we used a library of spectral templates created from
Sloan Digital Sky Survey data, covering spectral types M0--M9. For
each spectral type, we varied the flux contribution of the M-dwarf
template until the molecular absorption bands cancelled out as much as
possible in the difference spectrum of HS\,2214 minus template.  The best
match in the relative strength of the TiO absorption bands is achieved
for a spectral type M$2.5\pm0.5$, contributing 25\% of the observed
$V$-band flux of HS\,2214 (Fig.\,\ref{hs2214_secondary}).  The
extrapolated $JHK_\mathrm{s}$ spectrum of the secondary
star\footnote{Using LHS399 from Sandy Leggett's library of M-dwarf
spectra, http://ftp.jach.hawaii.edu/ukirt/skl/dM.spectra/} agrees
fairly well with the 2MASS $JHK_\mathrm{s}$ magnitudes of HS\,2214
(14.5, 13.9, and 13.5, respectively), suggesting that the accretion
disc contributes only a small amount to the infrared flux.

Using \citeauthor{beuermann+weichhold99-1}'s
(\citeyear{beuermann+weichhold99-1}) calibration of the surface
brightness in the 7165/7500\,\AA\ TiO band, and assuming a radius of
$R_2=(3.0\pm0.3)\times10^{10}$\,cm, based on the orbital period determined
below and various radius-orbital period relations
\citep[e.g.][]{warner95-1,beuermann+weichhold99-1}, we estimate the
distance of HS\,2214 to be $d=390\pm40$\,pc, where the error is dominated
by the uncertainty of the secondary's radius.

\subsection{The orbital period}

We first measured the radial velocity variation of \Ha\ in the INT
spectra and in the CAFOS spectra taken with the R-100 grism, as well
as that of \Hb\ in the B-100 CAFOS spectra by using the double
Gaussian method of \citet{schneider+young80-2}. The Scargle
periodogram calculated from these measurements contained a peak near
$\sim5.5$\,\id, but was overall of poor quality. In a second attempt,
we determined the \Ha\ and \Hb\ radial velocities by means of the
$V/R$ ratios, calculated from having equal fluxes in the blue and red
line wing, fixing the width of the line to $\sim2500$\,\kms\ in order
to avoid contamination by the \Line{He}{I}{6678} line adjacent
to \Ha. The Scargle periodogram calculated from these sets
of radial velocities contains the strongest signal at $6.6$\,\id\ and
an 1\,\id\ alias of similar strength at $5.6$\,\id\ (see
Fig.\,\ref{f-scargle}, bottom panel). Based on the spectroscopy alone,
an unambiguous period determination is not possible.

\begin{figure}
\centerline{\includegraphics[angle=-90,width=\columnwidth]{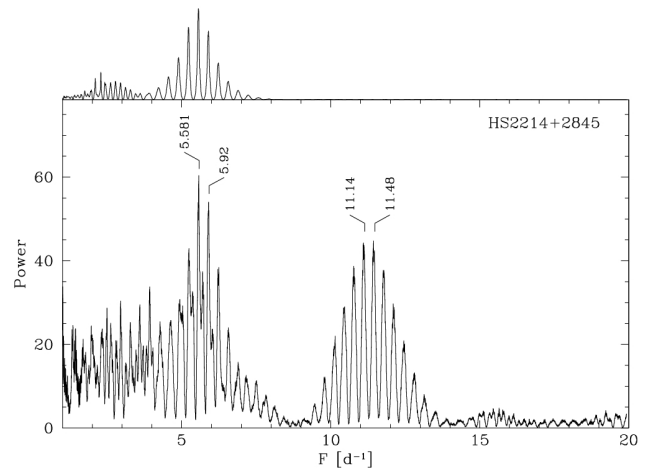}}
\caption [] {\label{f-aov_hs2214} Main panel: the analysis-of-variance
(AOV) periodogram of HS\,2214+2845 calculated from the two longest
light curves obtained at the Braeside Observatory, which show a
double-humped pattern. Top panel: the AOV periodogram created from a
sine wave with the orbital period of 258.02\,min.}
\end{figure}

\begin{figure}
\centerline{\includegraphics[width=\columnwidth]{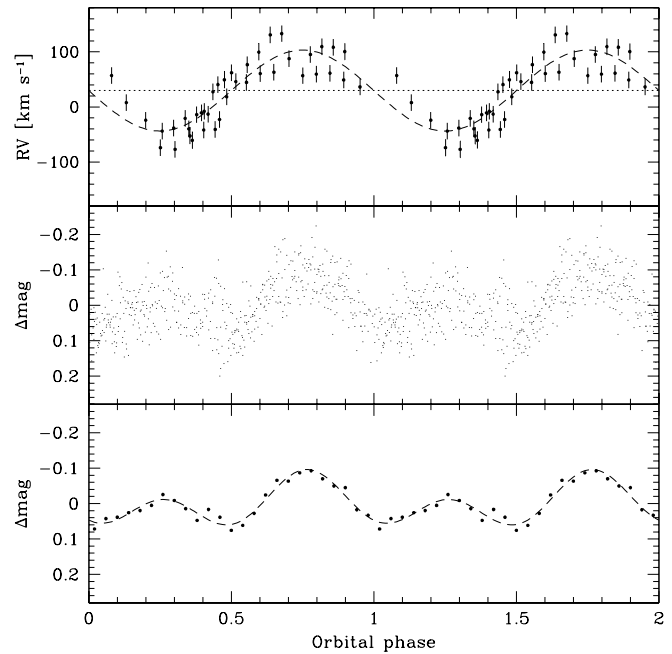}}
\caption [] {\label{f-fold_hs2214} The spectroscopic and
photometric data of HS\,2214 are folded on the photometric period of
258.02\,min using the spectroscopic zero-point
$T_{0}=\mathrm{HJD}\,2451812.3309$ defined by the red-to-blue crossing
of the \Ha\ and \Hb\ radial velocities. In that phase convention, the
inferior conjunction of the secondary star is expected at orbital
phase zero. Top panel: the radial velocity variation of the \Ha\ and
\Hb\ emission lines, as already shown in Fig.\,\ref{f-rvfolded}. A
whole cycle has been repeated for clarity. Middle panel: the
photometric data obtained from the Braeside Observatory. Bottom panel:
the photometric data binned into 25 phase slots, along with a
two-frequency sine fit (dashed line).}
\end{figure}

A crucial clue in determining the orbital period of HS\,2214 came from
the analysis of the two longest photometric time series obtained at
the Braeside Observatory in September 2000
(Fig.\,\ref{f-lc_hs2214}). These light curves display a double-humped
structure with a period of $\sim4$\,h, superimposed by relatively
low-amplitude flickering. The analysis-of-variance periodogram
{\citep[AOV,][]{schwarzenberg-czerny89-1}} calculated from these two
light curves contains two clusters of signals in the range of
$4-7$\,\id and $10-13$\,\id, respectively (see
Fig.\,\ref{f-aov_hs2214}). The strongest peaks in the first cluster
are found at $\simeq5.58$\,\id and $\simeq5.92$\,\id\, and at
$\simeq11.14$\,\id and $\simeq11.48$\,\id\ in the second
cluster. Based on the fact that two of the frequencies are
commensurate, we identify $f_1=5.58$\,\id\ and $f_2=11.14$\,\id\ as
the correct frequencies, with $f_1$ being the fundamental and $f_2$
its harmonic.  The periodogram of a faked data set computed from a
sine wave with a frequency of $5.58$\,\id, evaluated at the times of
the observations and offset by the randomised observational errors
reproduces the alias structure observed in the periodogram of the data
over the range $4-7$\,\id\ very well (Fig.\,\ref{f-aov_hs2214}, top
panel).  A two-frequency sine fit with $f_2=2\times f_1$ to the data
results in $f_1=5.581(12)$\,\id. The Braeside photometry folded over
that frequency displays a double-hump structure
(Fig.\,\ref{f-fold_hs2214}, two bottom panels). We identify $f_1$ as
the orbital frequency of the system, hence, $\Porb=258.02(56)$\,min
based on the following arguments. (a) The fundamental frequency
detected in the photometry coincides with that of the second-strongest
peak in the periodogram determined from the \Ha\ and \Hb\ $V/R$-ratio
radial velocity measurements (Fig.\,~\ref{f-scargle}, bottom
panel). (b) Double-humped orbital light curves are observed in a large
number of short-period dwarf novae, e.g.  WX\,Cet
\citep{rogoziecki+schwarzenberg-czerny01-1}, WZ\,Sge
\citep{patterson98-1}, RZ\,Leo, BC\,UMa, MM\,Hya, AO\,Oct, HV\,Vir
\citep{pattersonetal03-1}, HS\,2331+3905
\citep{araujo-betancoretal05-1}, and HS\,2219+1824
\citep{rodriguez-giletal05-1}; the origin of those double-humps is not
really understood, but most likely associated with the accretion
disc/bright spot. In long-period dwarf novae, double-humped light
curves are observed in the red part of the spectrum caused by
ellipsoidal modulation of the secondary star, e.g. U\,Gem
\citep{berrimanetal83-1} or IP\,Peg \citep{szkody+mateo86-1,
martinetal87-1}. In both cases, a strong and sometimes dominant,
signal at the harmonic of the orbital period is seen in the periodogram
calculated from their light curves.

\begin{figure}
\centerline{\includegraphics[angle=-90,width=\columnwidth]{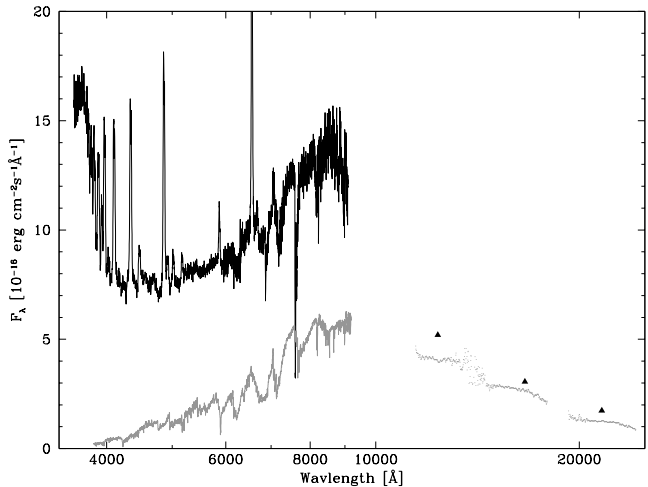}}
\caption[]{\label{hs2214_secondary} The average CAFOS spectrum of
  HS\,2214 (black line) along with the best-matching M-dwarf
  template of spectral type M2.5, scaled to fit the strength of the
  molecular bands in HS\,2214. The 2MASS $JHK_\mathrm{s}$ infrared
  fluxes of HS\,2214 are represented by filled triangles, the $JHK$
  spectrum of the M-dwarf template is shown as gray dots.}
\end{figure}

Figure\,\ref{f-rvfolded} (bottom panel) shows the radial velocity data
folded over the photometric orbital period (258.02\,min), along with a
sine-fit (Table\,\ref{t-rvfits}). The radial velocities are shown
again in Fig.\,\ref{f-fold_hs2214} together with the Braeside
photometry, all folded using the photometric period but the
spectroscopic zeropoint (Table\,\ref{t-rvfits}). The photometric minima
occur near orbital phase zero (inferior conjunction of the secondary)
and 0.5, consistent with what is expected for ellipsoidal modulation. Very
similar phasing is observed also for the double-humps in short-period
systems, e.g. WZ\,Sge \citep{patterson80-1} shows maximum brightness
close to phases 0.25 and 0.75. However, given the strong contribution
of the secondary star to optical flux of HS\,2214
(Fig.\,\ref{hs2214_secondary}), and the fact that the filterless
Braeside photometry is rather sensitive in the red, we believe that
the origin of the double-hump pattern seen in HS\,2214 is indeed
ellipsoidal modulation.

The binary parameters of HS\,2214 could be improved in a future study
by a measurement of the radial velocity of the secondary star, e.g. in
using the Na doublet the $I$ band, and a determination of the orbital
inclination from modelling the ellipsoidal modulation. 


\section{The orbital period distribution of dwarf novae\label{s-porbdn}}
Because of their outbursts, the vast majority of all currently known
dwarf novae have been discovered by variability surveys, either
through professional sky patrols, or through the concentrated efforts
of a large number of amateur astronomers
\citep{gaensicke05-1}. Considering the irregular temporal sampling of
such observations, the population of known dwarf novae is likely to be
biased towards systems which have frequent and/or large amplitude
outbursts.

%
\begin{figure*}
\includegraphics[angle=-90,width=\columnwidth]{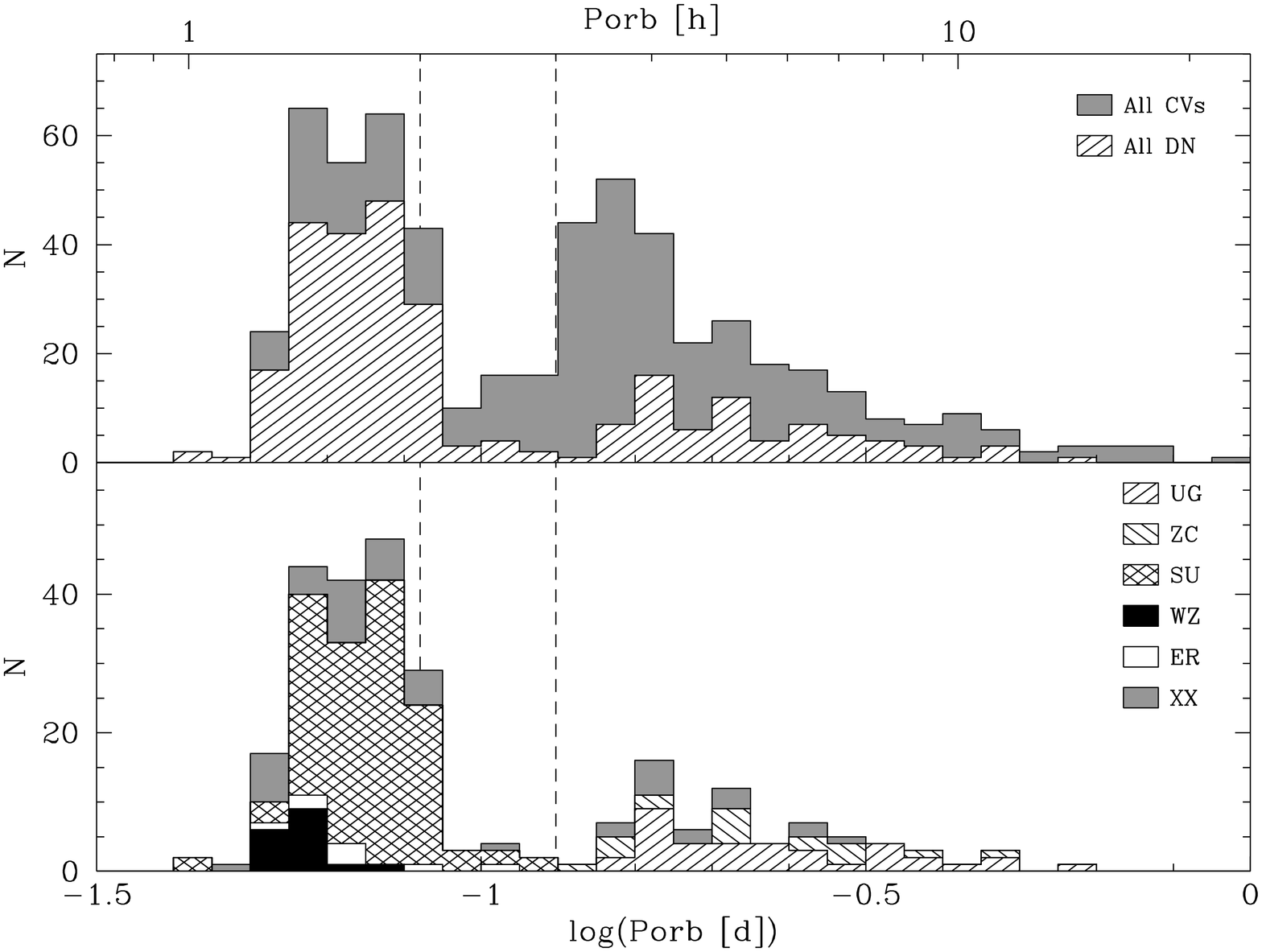}
\hfill
\includegraphics[angle=-90,width=\columnwidth]{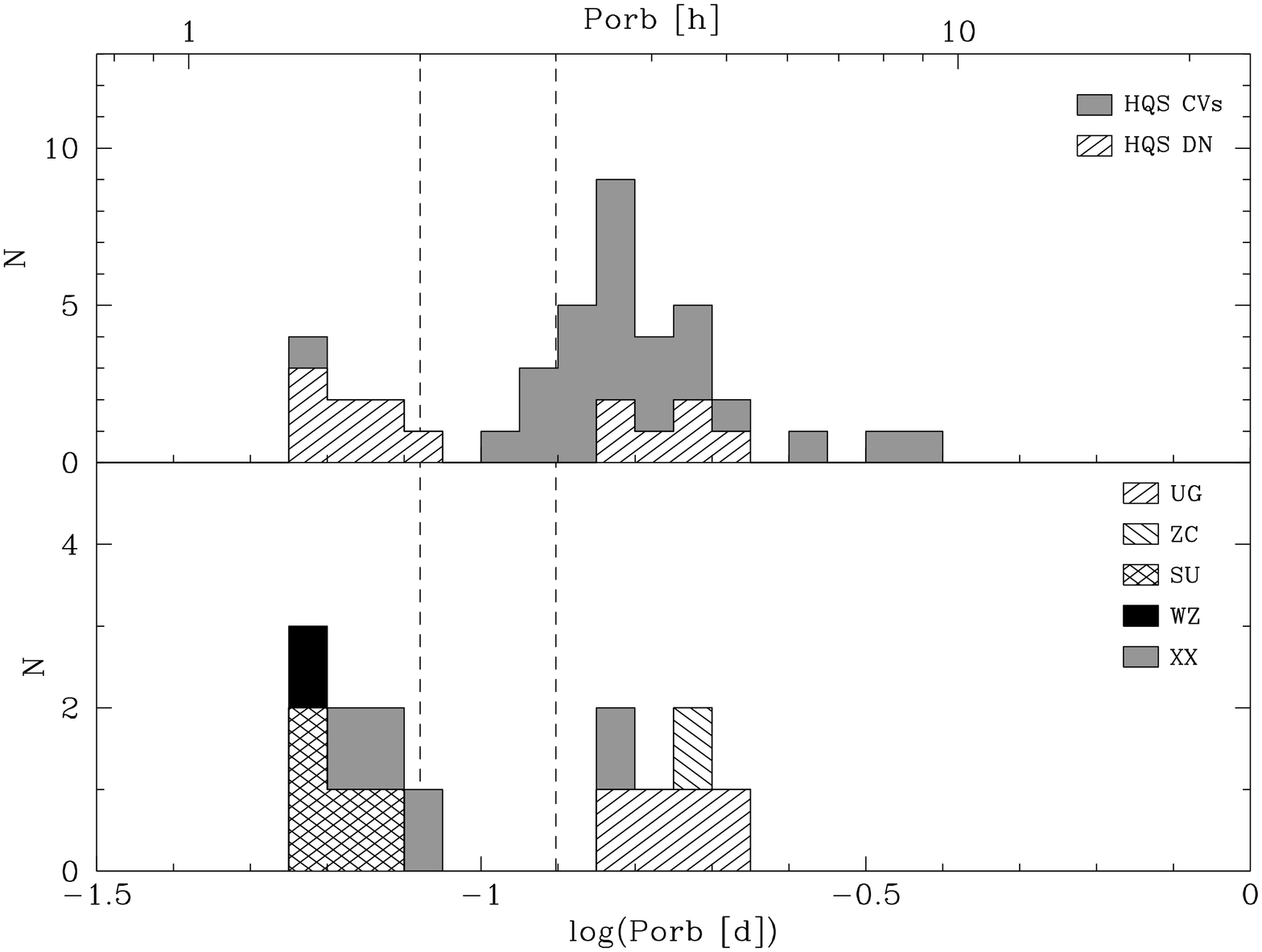}

\begin{minipage}[t]{\columnwidth}
\caption [] {\label{f-alldn_plot} Top panel: the orbital period
distribution of known CVs and dwarf novae from Ritter \& Kolb
(\citeyear{ritter+kolb03-1}, 7th Edition, rev. 7.5, July 2005) are shown
in gray and shade, respectively. Bottom panel: the period distribution
of known dwarf novae according to their subtype, U\,Gem (UG), Z\,Cam
(ZC), SU\,UMa (SU), WZ\,Sge (WZ), ER\,UMa (ER), and unclassified subtype
(XX). The dashed lines are the conventional 2--3\,h period gap.}
\end{minipage}
\hfill
\begin{minipage}[t]{\columnwidth}
\caption [] {\label{f-hqsdn_plot} Top panel: the orbital period
distribution of 41 new CVs and 14 dwarf novae identified in HQS are
shown in gray and shade, respectively. Bottom panel: the period
distribution of HQS dwarf novae according to their subtype, U\,Gem
(UG), Z\,Cam (ZC), SU\,UMa (SU), WZ\,Sge (WZ), and unclassified
subtype (XX). The dashed lines are the conventional 2--3\,h period gap.}
\end{minipage}
\end{figure*}

\begin{table} [t]
\caption[]{Dwarf novae discovered in the HQS with their
subtype, U\,Gem (UG), SU\,UMa (SU), Z\,Cam (ZC), and unclassified
(XX). Uncertain classifications are marked by a colon.\label{t-hqsdn}}
\setlength{\tabcolsep}{0.9ex}
\begin{tabular}{llccc}
\hline\noalign{\smallskip} 
HQS ID & Other name & \Porb\,(min) & Type & References\\
\hline\noalign{\smallskip}
HS\,2331+3905 &         & 81.1  & WZ: & 1 \\
HS\,1449+6415 & KV\,Dra & 84.9  & SU & 2,3 \\
HS\,2219+1824 &         & 86.3  & SU & 4 \\
HS\,1340+1524 &         & 92.7  & XX & 5 \\
HS\,1017+5319 & KS\,UMa & 97.9  & SU & 2,6 \\
HS\,0417+7445 &         & 105.1/109.9 & SU & 5 \\
HS\,1016+3412 &         & 114.3  & XX & 5 \\
HS\,0913+0913 & GZ\,Cnc & 127.1 & XX & 2,8,9 \\
HS\,0941+0411 & RX\,J0944.5+0357 & 215.0 & XX & 2,10 \\
HS\,0552+6753 & LU\,Cam & 216.6 & UG & 2,7 \\
HS\,0907+1902 & GY\,Cnc & 252.6 & UG & 2,11 \\
HS\,2214+2845 &         & 258.0 & UG & 5 \\
HS\,1857+7127 &         & 272.3 & ZC: & 5 \\
HS\,1804+6753 & EX\,Dra & 302.3 & UG & 12,13,14 \\
\noalign{\smallskip}\hline\noalign{\smallskip}
\end{tabular}
References: (1) \citet{araujo-betancoretal05-1}; (2)
\citet{jiangetal00-1}; (3) \citet{nogamietal00-1}; (4)
\citet{rodriguez-giletal05-1}; (5) this work; (6)
\citet{pattersonetal03-1}; (7) Thorstensen priv. com. \&
vsnet-campaign-dn 2681; (8) \citet{katoetal02-1}; (9)
\citet{tappert+bianchini03-2}; (10) \citet{mennickentetal02-1}; (11)
\citet{gaensickeetal00-2}; (12) \cite{fiedleretal97-1}; (13)
\cite{billingtonetal96-1}; (14) \citet{shafter+holland03-1}.
\end{table}

\subsection{The orbital period distribution of all known dwarf novae\label{s-alldn}}
Inspecting the \citeauthor{ritter+kolb03-1} catalogue
(\citeyear{ritter+kolb03-1}, Edition 7.5 of July 1, 2005)
within the orbital period range of $\sim1$\,h to $\sim1$\,d
and removing AM\,CVn systems, it is found that nearly half of all
known CVs (262 systems out of 572, or 46\%) are dwarf novae of which
166 (63\%) have $\Porb<2$\,h, 26 (10\%) are found in the $2-3$\,h
orbital period gap, and 70 (27\%) have long periods,
$\Porb>3$\,h. The conventional definition of the period gap as
being the range $2-3$\,h is somewhat arbitrary, and these numbers vary
slightly if a different definition is used, but without changing the
overall picture. Figure\,\ref{f-alldn_plot} (top panel) shows the
orbital period distribution of all known CVs and dwarf novae with
periods between $\sim1$\,h and $\sim1$\,d. Whereas the total
population of all CVs features the well-known period gap, i.e. the
relatively small number of CVs with periods $2$\,h$\,<\Porb<3$\,h, the
number of dwarf nova reaches a minimum in the range $3-4$\,h. In fact,
the number of dwarf nova with $3$\,h$\,<\Porb<4$\,h is a half (15, or
6\% of all dwarf novae) of that in the ``standard'' $2-3$\,h period
gap (26, or 10\%). The dearth of known dwarf novae in the $3-4$\,h
period range was pointed out by \citet{shafteretal86-1} and
\citet{shafter92-1}, who compared the observed dwarf nova period
distribution with those constructed from various magnetic braking
models, and concluded that none of the standard magnetic braking
models can satisfactorily explain the lack of observed dwarf novae in
the $3-4$\,h period range.

The bottom panel of Fig.\,\ref{f-alldn_plot} displays all known dwarf
novae according to their subtypes which are 159 (61\%) SU\,UMa, 37
(14\%) U\,Gem, 18 (7\%) Z\,Cam, and 48 (18\%) unclassified subtypes
(XX). For completeness, we note that the SU\,UMa class includes 8
ER\,UMa stars (which have very short superoutburst cycles) and 19
WZ\,Sge stars (which have extremely long outburst cycles). All
confirmed U\,Gem and Z\,Cam stars lie above the period gap, in fact
all but one Z\,Cam star (BX\,Pup) have
$\Porb>3.5$\,h\footnote{\citet{ritter+kolb03-1} list five U\, Gem-type
dwarf novae with $\Porb<3$\,h: CC\,Cnc is a SU\, UMa-type dwarf nova
\citep{kato+nogami97-1}, and we included 587\,Lyr, CF\,Gru, V544\,Her,
and FS\,Aur as dwarf novae with no subtype (XX) due to the lack of
clear observational evidence for a specific subtype.}.  It is clearly
seen that the majority (85\%) of SU\,UMa lie below the period gap and
only a small fraction (15\%) inhabits the $2-3$\,h period
range\footnote{A note of caution among the WZ\,Sge stars, which mostly
have ultrashort-periods, concerns UZ\,Boo. \citet{ritter+kolb03-1}
list a period of $\sim3$\,h based on quiescent photometry, which is
almost certainly wrong. Intensive time-series of the 2003 outburst of
UZ\,Boo revealed a superhump period of 89.3\,min (Kato,
vsnet-campaign-dn 4064), and we use here this value as an estimate of
the orbital period.}.

The orbital period distribution of short-period dwarf novae in
Fig.\,\ref{f-alldn_plot} (the majority of all CVs in this period
range) differs markedly from the predictions made by the standard CV
evolution theory \citep[e.g.][]{kolb93-1, howelletal01-1,
kolb+baraffe99-1}: the minimum period is close to $\sim77$\,min,
contrasting with the predicted minimum period of $\sim65$\,min
\citep{paczynski+sienkiewicz83-1}, and the distribution of systems is
nearly flat in \Porb, whereas the theory predicts a substantial
accumulation of systems at the minimum period. Several modifications
of the CV evolution theory have been suggested to resolve this
discrepancy, however, none with undisputable success \citep{kingetal02-1,
renvoizeetal02-1, barker+kolb03-1}.

\subsection{The orbital period distribution of dwarf novae in the HQS\label{s-hqsdn}}
Another possible explanation for the lack of a spike in the orbital
period distribution of CVs near the minimum period is that systems
close to the minimum period, especially those evolving back to longer
periods, have not yet been discovered due to observational selection
effects. As most CVs below the period gap are dwarf novae, the most
obvious bias suppressing the period spike is to assume that dwarf
novae close to the minimum period have very rare outbursts. In fact, a
number of dwarf novae near the minimum period have very long outburst
intervals, e.g. WZ\,Sge ($\Porb=81.6$\,min, \citealt{pattersonetal02-2})
erupts every $20-30$ years; GW\,Lib ($\Porb=76.8$\,min,
\citealt{thorstensenetal02-3}) has been seen in outburst only once in
1983. It can not be excluded that these systems represent ``the tip of
the iceberg'' of a dwarf nova population with even longer outburst
periods. Assuming that rarely outbursting dwarf novae do exist in a
significant number, and that they spectroscopically resemble the known
objects, such as WZ\,Sge or GW\,Lib, our search for CVs in the HQS
should be able to identify them \citep{gaensickeetal02-2}.

We have so far obtained orbital periods for 41 new CVs found in the
HQS, their period distribution is shown in Fig.\,\ref{f-hqsdn_plot}
(top panel). The first thing to notice is that the majority of the new
CVs identified in the HQS are found \textit{above} the period gap, with a
large number of systems in the period range $3-4$\,h \citep[for a
discussion of the properties of CVs in this period range,
see][]{aungwerojwitetal05-1}. As for the overall CV population, a
dearth of systems is observed in the $2-3$\,h period range
(Fig.\,\ref{f-hqsdn_plot}, top panel), with the gap being wider for
dwarf novae (Fig.\,\ref{f-hqsdn_plot}, bottom panel).

To date 14 (26\%) out of 53 new CVs discovered in the HQS have been
classified as dwarf novae, including the five systems, HS\,0417,
HS\,1016, HS\,1340, HS\,1857, and HS\,2214, presented in this paper
(Table \ref{t-hqsdn}). The fraction of long-period ($\Porb>3$\,h)
systems is larger in the HQS sample (43\%) than in the total
population of known dwarf novae (27\%, see Sect.\,\ref{s-alldn}). 
The total number of new HQS dwarf novae is relatively small,
and subject to corresponding statistical uncertainties. However, the
tilt towards long-period dwarf novae among the new HQS CVs is likely
to be underestimated, as a significant number of long-period HQS CVs
have still uncertain CV subtypes, and several of them could turn out
to be additional Z\,Cam-type dwarf nova
(\citealt{aungwerojwitetal05-1} plus additional unpublished
data). Optical monitoring of the long-term variability of these
systems will be necessary to unambiguously determine their CV type.
Overall, the dwarf novae identified within the HQS fulfill the above
expectations of being ``low-activity'' systems, i.e. dwarf novae that
have either infrequent outbursts (e.g. KV\,Dra, HS\,0941+0411,
HS\,2219+1824) or low-amplitude outbursts (e.g. EX\,Dra). We found
only one system that resembles the WZ\,Sge stars with their very long
outburst intervals found near the minimum period that is HS\,2331+3905
\citep{araujo-betancoretal05-1} which has a period of 81.1\,min, and no
outburst has been detected so far. 

Thus, our search for CVs in the HQS has been unsuccessful in
identifying the predicted large number of short-period CVs, despite
having a very high efficiency in picking up systems that resemble the
typical known short-period dwarf novae.

\section{Constraints on the space density of CVs\label{s-spacedensity}}
CV population models result in space densities in the range
$2\times10^{-5}\,\mathrm{pc}^{-3}$ to
$2\times10^{-4}\,\mathrm{pc}^{-3}$ \citep{dekool92-1, politano96-1},
whereas the space density determined from observations is
$(0.5-1)\times10^{-5}\,\mathrm{pc}^{-3}$ \citep{patterson84-1,
ringwald96-1, patterson98-1}. It appears therefore that we currently
know about an order of magnitude less CVs than predicted by the
models. Also the observed ratio of short to long orbital period
systems is in strong disagreement with the predictions of the
theory. \citet{patterson84-1, patterson98-1} estimated that the
fraction of short-period CVs per volume is $75-80$\%, which has to be
compared to 99\% in the population studies
\citep{kolb93-1,howelletal97-1}.

Because of the large differences in mass transfer rates, and, hence,
absolute magnitudes, of long ($\Porb>3$\,h) and short ($\Porb<3$\,h)
period CVs, magnitude limited samples appear at a first glance utterly
inappropriate for the discussion of CV space densities. However,
taking the theoretical models at face value, the space density of long
period CVs is entirely negligible compared to that of short period CVs
\citep{kolb93-1, howelletal97-1}, and hence a discussion of the total
CV space density can be carried out on the basis of short-period
systems alone. In the following, we assess the expected numbers of
systems in the HQS separately for short-period CVs that are still
evolving towards the minimum period (pre-bounce), and those that
already reached the minimum period, and evolve back to longer periods
(post-bounce). For both cases, we assumed (1) a space density of
$5\times10^{-5}\,\mathrm{pc}^{-3}$ as an intermediate value between
the predictions of \citet{dekool92-1} and \citet{politano96-1}, (2)
that 70\% of all CVs are post-bounce systems, and 30\% are pre-bounce
systems \citep{kolb93-1, howelletal97-1}(ignoring, as stated above,
the small number of long-period CVs), (3) a scale height of 150\,pc
\citep[e.g.][]{patterson84-1}, and (4) that the luminosity of
short-period CVs is dominated by the accretion-heated white dwarf.

\subsection{Pre-bounce CVs expected in the HQS}
CV evolution models predict a typical accretion rate of $\dot
M\simeq5\times10^{-11}\,\msy$ for pre-bounce systems, with a
relatively small spread \citep{kolb93-1, howelletal97-1}. For this
accretion rate, \citeauthor{townsley+bildsten03-1}'s (2003)
calculation of white dwarf accretion heating predicts an effective
temperature of $T_{\mathrm{wd}}\simeq12\,000$\,K. If the luminosity is
dominated by a white dwarf of this temperature, the HQS with an
average magnitude limit of $B\simeq18$ would detect pre-bounce CVs out
to a distance of 175\,pc (using the absolute magnitudes of white
dwarfs by \citealt{bergeronetal95-2}), i.e. over a bit more than one scale
height. Within a sphere of radius 175\,pc around the Earth, one would
expect $\simeq144$ CVs (taking into account the exponential drop-off
of systems perpendicular to the plane), of which $\simeq36$ would be
within the sky area sampled by the HQS. This is a  conservative
lower limit, as any additional luminosity from the accretion disc
and/or bright spot as well as a hotter white dwarf temperature would
increase the volume sampled by the HQS, and therefore increase the
number of pre-bounce CVs within the survey.

\subsection{Post-bounce CVs expected in the HQS}
Once past the minimum period, the accretion rate of CVs substantially
drops as a function of time, and we assume here a value of $\dot
M=10^{-11}\,\msy$, corresponding to an intermediate CV age of
$\sim5$\,Gyr \citep{kolb+baraffe99-1}, and  a white dwarf
temperature of $T_\mathrm{wd}\simeq7000$\,K
\citep{townsley+bildsten03-1}. This lower white dwarf temperature
reduces the detection limit of the HQS to only $\simeq65$\,pc. The
total number of post-bounce CVs within a sphere of radius 65\,pc
around the Earth is hence expected to be 40, of which 10 within the
HQS area.

\subsection{Short-period CVs in the HQS: most likely pre-bounce only}
The immediate question is \textit{what type are the short-period HQS
CVs: pre- or post-bounce?}  As mentioned in Sect.\,\ref{s-hqsdn}, only
12 short-period ($\Porb<3$\,h) systems have been found among the 41
new HQS CVs for which we have adequate follow-up data. The new
short-period CVs comprise 8 dwarf novae (Table\,~\ref{t-hqsdn}),
two polars \citep{reimersetal99-1, jiangetal00-1, schwarzetal01-1,
tovmassianetal01-2, thorstensen+fenton02-1, gaensickeetal04-3}, one
intermediate polar \citep{rodriguez-giletal04-1, pattersonetal04-1},
and one system with uncertain classification
\citep{gaensickeetal04-1}. The white dwarf has been detected
in the spectra of HS\,2237+8154 \citep{gaensickeetal04-1},
HS\,2331+3905 \citep{araujo-betancoretal05-1}, HS\,2219+1824
\citep{rodriguez-giletal05-1}, and HS\,1552+2730
\citep{gaensickeetal04-3}, with temperatures of $\simeq10\,500$\,K,
$\simeq11\,500$\,K, $\simeq15\,000$\,K, and $\simeq20\,000$\,K,
respectively. These four systems are very likely to have the lowest
mass transfer rates among the 12 new short-period CVs, the optical
spectra of the other eight are characterised by strong Balmer and He
emission lines and the associated continuum which outshines the white
dwarf, typical of higher accretion rates.  While there are still about
a dozen HQS CVs with no accurate orbital period determination, the
data already at hand makes it very unlikely that more than 2 or 3 of
those systems will turn out to have periods $<3$\,h.

For a complete assessment of the short-period content of the HQS, one
has obviously to include in the statistics the short-period CVs that
are contained within the HQS data base, but were already
known~--~subject to the same selection criteria that were applied to
identify the twelve new systems. \citet{gaensickeetal02-2} analysed
the properties of the previously known CVs within the HQS data base,
and came to the following conclusions.  18 previously known
short-period ($\Porb<3$\,h) systems with HQS spectra are correctly
(re-)identified as CVs, including 12 dwarf novae, five polars, and one
intermediate polar\footnote{Excluding the double-degenerate helium
CVs, which follow a different evolution channel that is not taken into
account in the population models of \citet{politano96-1} and
\citet{dekool92-1}.}. \citet{gaensickeetal02-2} also found that only
two previously known short-period systems with HQS spectra failed to
be identified as CVs; this ``hit rate'' of 90\% underlines the extreme
efficiency of the HQS of finding short period CVs.  Five out of those
18 systems have measured white dwarf temperatures, all of them in the
range $\simeq11\,000-16\,000$\,K (MR\,Ser, ST\,LMi, AR\,UMa, SW\,UMa,
T\,Leo: \citealt{gaensickeetal01-1, hamilton+sion04-1,
araujo-betancoretal05-2, gaensickeetal05-2}). The remaining 13
systems all have spectra dominated by strong Balmer and He emission,
suggesting accretion rates too high to detect the white dwarf.

In summary, the HQS contains a total of 30 short-period CVs (12 new
identifications plus 18 previously known systems), all of which are 
consistent with being pre-bounce systems. At face value, this  
number agrees rather nicely with the 36 expected systems derived
above, but one has to bear in mind that that number is an absolute
lower limit, as hotter white dwarfs and/or accretion luminosity from
the disc and hot spot will increase the volume sampled by the HQS. 

While there may still be some shortfall of pre-bounce systems, it is
much more worrying that so far no systems with the clear signature of
a post-bounce CV that evolved significantly back to longer periods has
been found~--~neither in the HQS, nor elsewhere. The coldest CV white
dwarfs have been found, to our knowledge, in the polar EF\,Eri
($T_\mathrm{wd}\simeq9500$\,K, \citealt{beuermannetal00-1}), and
HS\,2331+3905 ($T_\mathrm{wd}\simeq10\,500$\,K,
\citealt{araujo-betancoretal05-1}), both systems with orbital periods of
$\simeq81$\,min~--~which may hence be either pre- or post-bounce
systems. 

A final note concerns the number of WZ\,Sge stars, i.e. short-period
dwarf novae with extremely long outburst intervals. Given the strong
Balmer lines in the known WZ\,Sge stars e.g. WZ\,Sge itself
\citep{gillilandetal86-1}, BW\,Scl \citep{abbottetal97-1}, GD\,552
\citep{hessman+hopp90-1}, and GW\,Lib
\citep{szkodyetal00-1,thorstensenetal02-3}, we believe that any
WZ\,Sge brighter than $B\simeq18$ would have easily been identified in the
HQS~--~yet, only a single new WZ\,Sge system has been discovered,
HS\,2331+3905 \citep{araujo-betancoretal05-1}.

Thus, we conclude that while our systematic effort in identifying new
CVs leads to a space density of pre-bounce short period CVs which
agrees with the predictions within an order of magnitude, the bulk of
all CVs, which are predicted to have made it past the minimum orbital
period, remains unidentified so far.

\section{Conclusions}
We have identified five new dwarf novae as part of our search for new
CVs in the HQS, bringing the total number of HQS-discovered dwarf
novae to 14. The new systems span orbital periods from $\sim1.5$\,h to
nearly 5\,h, confirming the trend that dwarf novae spectroscopically
selected in the HQS display a larger ratio of long-to-short orbital
periods. Overall, dwarf novae represent only about one third of the
HQS CVs which are studied sufficiently well, and it is by now clear
that the properties of the sample of new HQS CVs do not agree with
those of the predicted large population of short-period
low-mass-transfer systems.  Within the limiting magnitude of
$B\simeq17.5-18.5$, the HQS \textit{does} however contain a large
number of previously known dwarf novae that were identified because of
their variability, and almost all of these systems have been recovered
as strong CV candidates on the basis of their HQS spectrum
\citep{gaensickeetal02-2}. Based on their spectroscopic and
photometric properties it appears that most, if not all short-period
CVs in the HQS (newly identified and previously known) are still
evolving towards the minimum period. If the large number of
post-bounce CVs evolving back to longer periods predicted by
population models exists, they must (a) have very long outburst
recurrence times, and (b) have H$\beta$ equivalent widths that are far
lower than observed in the currently known typical short-period CVs.

\acknowledgements AA thanks the Royal Thai Government for a
studentship. BTG and PRG were supported by a PPARC Advanced Fellowship
and a PDRA grant, respectively. MAPT is supported by NASA LTSA grant
NAG-5-10889. RS is supported by the Deutsches Zentrum f\"ur Luft und
Raumfahrt (DLR) GmbH under contract No.\,FKZ \mbox{50 OR 0404}.  AS is
supported by the Deutsche Forschungsgemeinschaft through grant
Schw536/20-1. The HQS was supported by the Deutsche
Forschungsgemeinschaft through grants Re\,353/11 and Re\,353/22. We
thank Tanya Urrutia for carrying out a part of the AIP
observations. PS thanks Robert Mutel (University of Iowa) and his
students for taking CCD images with the Rigel telescope. Tom Marsh is
acknowledged for developing and sharing his reduction and analysis
package \texttt{MOLLY}. This publication makes use of data products
from the Two Micron All Sky Survey, which is a joint project of the
University of Massachusetts and the Infrared Processing and Analysis
Center/California Institute of Technology, funded by the National
Aeronautics and Space Administration and the National Science
Foundation.

Based in part on observations collected at the Centro Astron\'omico
Hispano Alem\'an (CAHA) at Calar Alto, operated jointly by the
Max-Planck Institut f\"ur Astronomie and the Instituto de
Astrof{\'\i}sica de Andaluc{\'\i}a (CSIC); on observations made at the
1.2m telescope, located at Kryoneri Korinthias, and owned by the
National Observatory of Athens, Greece; on observations made with the
IAC80 telescope, operated on the island of Tenerife by the Instituto
de Astrof{\'\i}sica de Canarias (IAC) at the Spanish Observatorio del
Teide; on observations made with the OGS telescope, operated on the
island of Tenerife by the European Space Agency, in the Spanish
Observatorio del Teide of the IAC; on observations made with the Isaac
Newton Telescope, which is operated on the island of La Palma by the
Isaac Newton Group in the Spanish Observatorio del Roque de los
Muchachos of the IAC; on observations made at the Wendelstein
Observatory, operated by the Universit\"ats-Sternwarte M\"unchen; on
observations made with the 1.2m telescope at the Fred Lawrence Whipple
Observatory, a facility of the Smithsonian Institution; and on
observations made with the NASA/ESA Hubble Space Telescope, obtained
at the Space Telescope Science Institute, which is operated by the
Association of Universities for Research in Astronomy, Inc., under
NASA contract NAS 5-26555.

\bibliographystyle{aa}

\end{document}